\documentclass[12pt]{amsart}
\usepackage{amsmath,amssymb,cite,enumerate,graphicx,hyperref,morefloats}

\hypersetup{pdftex,colorlinks=true,allcolors=blue}
\usepackage[all]{hypcap}

\usepackage[foot]{amsaddr}
\usepackage[sc]{mathpazo}

\newcommand\ack{\subsection*{Acknowledgment}}

\DeclareMathAlphabet\mathsfbi{T1}{phv}{b}{it}

\numberwithin{equation}{section}

\newcommand\BV{\boldsymbol} 
\newcommand\BM{\mathsfbi} 

\newcommand\dif{\,\mathrm{d}}

\newcommand\parderiv[2]{\frac{\partial #1}{\partial #2}}

\renewcommand\div{\mathrm{div}}

\newcommand\trace{\mathrm{tr}}

\newcommand\Pran{\mbox{\textit{Pr}}}

\begin{document}

\author[Rafail V. Abramov]{Rafail V. Abramov}

\address{Department of Mathematics, Statistics and Computer Science,
University of Illinois at Chicago, 851 S. Morgan st., Chicago, IL 60607}

\email{abramov@uic.edu}

\author[Jasmine T. Otto]{Jasmine T. Otto}

\email{jotto3@uic.edu}

\date{\today}

\title{Nonequilibrium diffusive gas dynamics: Poiseuille microflow}

\begin{abstract}
We test the recently developed hierarchy of diffusive moment closures
for gas dynamics together with the near-wall viscosity scaling on the
Poiseuille flow of argon and nitrogen in a one micrometer wide
channel, and compare it against the corresponding Direct Simulation
Monte Carlo computations. We find that the diffusive regularized Grad
equations with viscosity scaling provide the most accurate
approximation to the benchmark DSMC results. At the same time, the
conventional Navier-Stokes equations without the near-wall viscosity
scaling are found to be the least accurate among the tested closures.
\end{abstract}


\maketitle

\section{Introduction}

The diffusive Boltzmann equation and the corresponding hierarchy of
the diffusive moment closure equations were derived in a recent
work~\cite{Abr13}, including the equilibrium (diffusive Navier-Stokes)
and nonequilibrium (diffusive Grad) closures. The additional
mass-diffusive term originated from the difference between the
deterministic real gas dynamics and the artificial ``random gas''
multimolecular process, which led to the conventional Boltzmann
equation. Additionally, a near-wall viscosity scaling was computed
in~\cite{Abr15}, based on the shortened mean free path of a gas
molecule near a wall. The new equations were studied in a Couette
microflow setting for argon and nitrogen, and it was found that the
diffusive Navier-Stokes and Grad moment closures with the near-wall
viscosity scaling developed Knudsen velocity boundary layers near the
walls, closely matching the results of the corresponding DSMC
computations. It was also found that the component of the heat flux
parallel to the flow, produced by the DSMC computations, was captured
quite well by the diffusive regularized Grad equations, but not by the
conventional or diffusive Navier-Stokes closures.

In the present work, we test the developed hierarchy of diffusive
moment closures~\cite{Abr13} together with the near-wall viscosity
scaling~\cite{Abr15} on the Poiseuille flow of argon and nitrogen in a
one micrometer wide channel, and compare it against the corresponding
Direct Simulation Monte Carlo (DSMC)
computations~\cite{Bird,ScaRooWhiDarRee}.  We find that the diffusive
regularized Grad equations with viscosity scaling provide the most
accurate approximation to the benchmark DSMC results. At the same
time, the conventional Navier-Stokes equations are found to be the
least accurate among the tested closures.

The paper is organized as follows. In Section~\ref{sec:theory} we
present the diffusive closure equations~\cite{Abr13} and near-wall
viscosity scaling~\cite{Abr15}. In Section~\ref{sec:experiments} we
show the results of the computational experiments with the Poiseuille
flow. In Section~\ref{sec:summary} we summarize the results of the
work.

\section{The diffusive closures for continuum gas dynamics}
\label{sec:theory}

The diffusive equations for the density $\rho$, velocity $\BV u$ and
energy $E$ are given by
\begin{subequations}
\label{eq:rho_u_E}
\begin{equation}
\parderiv\rho t+\div(\rho\BV u)=\div\left(\frac{D_\alpha}p\nabla p
\right),
\end{equation}
\begin{equation}
\parderiv{(\rho\BV u)}t+\div(\rho(\BV u\BV u^T+\BM T))=\div
\left(\frac{D_\alpha}p\nabla (p\BV u)\right),
\end{equation}
\begin{equation}
\parderiv{(\rho E)}t+\div(\rho(E\BV u+\BM T\BV u+\BV q))=\div
\left(\frac{D_\alpha}p\nabla (pE)\right),
\end{equation}
\end{subequations}
where $D_\alpha$ is the empirically scaled mass diffusivity of the
gas~\cite{Abr13}. Above, the pressure $p$ and the temperature tensor
$\BM T$ are further given by
\begin{equation}
\label{eq:p_theta}
p=\rho\theta,\qquad\BM T=\theta\BM I+\BM S,\qquad\theta=(\gamma-1) \left(
E-\frac 12\|\BV u\|^2\right),
\end{equation}
with $\theta$ and $\gamma$ being the temperature of the gas in energy
units and the adiabatic exponent of the gas, respectively. As we can
see, the equations in \eqref{eq:rho_u_E}--\eqref{eq:p_theta} are
closed under all variables except for the stress $\BM S$ and heat flux
$\BV q$.

\subsection{The Navier-Stokes closure for the stress and heat flux}

In the Navier-Stokes closure of the gas dynamics, the stress $\BM S$
and heat flux $\BV q$ are approximated from the Newton and Fourier
laws as follows~\cite{Abr13}:
\begin{subequations}
\label{eq:S_q_NS}
\begin{equation}
\rho\BM S_{NS}=-\mu\left(\nabla\BV u+(\nabla\BV u)^T+(1-\gamma)(\div\BV
u)\BM I\right),
\end{equation}
\begin{equation}
\rho\BV q_{NS}=-\frac\gamma{\gamma-1}\frac\mu\Pran\nabla\theta.
\end{equation}
\end{subequations}
Above, $\mu$ and $\Pran$ are the viscosity and the Prandtl number of
the gas, respectively. Together with~\eqref{eq:rho_u_E}
and~\eqref{eq:p_theta}, the relations in~\eqref{eq:S_q_NS} comprise
the diffusive Navier-Stokes equations~\cite{Abr13}. The conventional
Navier-Stokes equations~\cite{Bat} result by setting the scaled mass
diffusivity coefficient $D_\alpha$ to zero.

\subsection{The Grad equations for the stress and heat flux}

In the diffusive Grad closure of the gas dynamics, the stress $\BM S$
and heat flux $\BV q$ are endowed with their own transport
equations~\cite{Abr13,Gra,Mal}
\begin{subequations}
\label{eq:S_q}
\begin{multline}
\parderiv{(\rho\BM S)}t+\div(\rho(\BV u\otimes\BM S+\BM Q))+\left(\BM
P+\BM P^T+(1-\gamma)\trace(\BM P)\BM I\right)=\\=-\frac{\rho p}\mu\BM
S+\div\left(\frac{D_\alpha}p \nabla(p\BM S)\right),
\end{multline}
\begin{multline}
\parderiv{(\rho\BV q)}t+\div(\rho\BV u\BV q)+\div(\BM P_2\BM T_2)-\BM
T_2\div\BM P_2-\div(\rho\BM S^2)+\\+\rho(\nabla\BV u)^T\BV q+\frac{
  \gamma-1}\gamma\rho\left[\nabla\BV u+(\nabla\BV u)^T+(\div\BV u)\BM
  I\right]\BV q+\\+\rho\BM Q:(\nabla\BV u)+\div(\rho\BM R)=-\Pran
\frac{\rho^2\theta}\mu\BV q +\div\left(\frac {D_\alpha}p\nabla(p\BV q)
\right),
\end{multline}
\end{subequations}
where
\begin{subequations}
\begin{equation}
\BM P=\left(\rho\BM T-D_\alpha(\nabla\BV u)^T\right)\nabla\BV u+
\frac{\gamma-1}\gamma\nabla(\rho\BV q),
\end{equation}
\begin{equation}
\BM T_2=\BM T+\frac\theta{\gamma-1}\BM I, \qquad\BM P_2=\rho\BM
T-2D_\alpha\nabla\otimes\BV u.
\end{equation}
\end{subequations}
Observe that the equations in~\eqref{eq:rho_u_E} and~\eqref{eq:S_q}
are not closed with respect to the matrix $\BM R$ and the 3-rank
tensor $\BM Q$ above.  For the classical Grad closure~\cite{Gra,Gra2},
both $\BM Q$ and $\BM R$ are set to zero. For the regularized Grad
closure~\cite{Abr13,Stru,StruTor,TorStru}, $\BM Q$ and $\BM R$ are set
to
\begin{subequations}
\label{eq:Reg_Grad}
\begin{equation}
\BM Q=\widetilde{\!\BM Q}+\widetilde{\!\BM Q}^T+\widetilde{\!\BM
  Q}^{TT},
\end{equation}
\begin{equation}
\BM R=\widetilde{\!\BM R}+\widetilde{\!\BM R }^T+\Big(\widetilde
R+(1-\gamma)\trace(\widetilde{\!\BM R})\Big)\BM I,
\end{equation}
\end{subequations}
where the notations $\widetilde{\!\BM Q}$, $\widetilde R$ and
$\widetilde{\!\BM R}$ read
\begin{subequations}
\label{eq:Reg_Grad_2}
\begin{multline}
\rho\,\widetilde{\!\BM Q}=-\frac\mu{\Pran_{\widetilde{\!\BM Q}}}\bigg[
  \nabla\BM S-\frac{\gamma-1}\gamma\BM I\otimes\div\BM S-\frac 1p
  \left(\BM S\otimes\div(\rho\BM S)-\frac{\gamma-1}\gamma\BM I\otimes
  \BM S\,\div(\rho\BM S)\right)+\\+\frac{\gamma-1}{\gamma\theta}
  \bigg(\BV q\otimes\left(\nabla\BV u+(\nabla\BV u)^T\right)-\frac{
    \gamma-1}\gamma\BM I\otimes\left(\nabla\BV u+(\nabla\BV u)^T+(
  \div\BV u)\BM I\right)\BV q\bigg)\bigg],
\end{multline}
\begin{equation}
\rho\widetilde R=-\frac{2\mu}{\Pran_{\widetilde R}}\left[\frac\gamma
  \theta\div(\theta\BV q)-\div\BV q+(\gamma-1)\left(\BM S:(\nabla\BV
  u) -\frac 1p\BV q^T\div(\rho\BM S)\right) \right],
\end{equation}
\begin{multline}
\rho\,\widetilde{\!\BM R}=-\frac\mu{\Pran_{\widetilde{\!\BM R}}}\bigg[
  \BM S\left(\nabla\BV u+(\nabla\BV u)^T\right)+\frac{2\gamma-1}{
    \gamma\theta}\left(\nabla(\theta\BV q)-\frac 1\rho\BV q
  \div(\rho\BM S)^T\right)-\\-\left((\gamma-1)\div\BV u+\frac{
    2\gamma-1}{2\theta}\left(\frac 1\rho\div(\rho\BV q)+\BM
  S:(\nabla\BV u)\right)\right)\BM S\bigg].
\end{multline}
\end{subequations}
Above, the constants $\Pran_{\widetilde{\!\BM Q}}$, $\Pran_{\widetilde
  R}$ and $\Pran_{\widetilde{\!\BM R}}$ are the third- and
fourth-moment Prandtl numbers, which equal $3/2$, $2/3$ and $7/6$,
respectively, for an ideal monatomic gas \cite{Stru,StruTor,TorStru}.

\subsection{Near-wall scaling of viscosity and mass diffusivity}

Previously in~\cite{Abr13,Abr15}, it was found that, due to the
shrinkage of the molecular mean free path in the vicinity of a wall,
the viscosity and mass diffusivity should be scaled near channel walls
as
\begin{equation}
\label{eq:viscosity_scaling}
\frac{\mu^{\text{near wall}}}\mu=\frac{D_\alpha^{\text{near wall}}}{
  D_\alpha} =1+\frac 12\left(\frac x\lambda E_1(x/\lambda)
-e^{-x/\lambda}\right),
\end{equation}
where $x$ is the distance to the wall, $\lambda$ is the standard
length of the mean free path away from the wall, and $E_1(x)$ is the
exponential integral:
\begin{equation}
E_1(x)=\int_x^\infty\frac{e^{-y}}y\dif y.
\end{equation}
In the current work, we use the viscosity and mass diffusivity scaling
above for all studied closures. We also compare the results against
the conventional Navier-Stokes closure without the viscosity scaling.
As in~\cite{Abr13}, for the computation of $E_1(x)$ we use the
approximation proposed in \cite{SwaOjh}. To estimate the mean free
path $\lambda$ from the thermodynamic quantities, we use the
approximate formula given in \cite{Cer2}, Chapter 5, eq. (1.3):
\begin{equation}
\label{eq:mfp}
\lambda=\frac\mu p\sqrt{\frac{\pi\theta}2}.
\end{equation}

\section{Poiseuille microflow}
\label{sec:experiments}

\subsection{The DSMC set-up}

The Direct Simulation Monte Carlo (DSMC) method \cite{Bird} models a
gas flow in a direct fashion by computing the motion and collisions of
the actual gas molecules. Previously, we used the DSMC simulations
in~\cite{Abr13,Abr15} to model the Couette flow in a microchannel.
For the DSMC simulation of the Poiseuille flow in the present work we
used the dsmcFoam\footnote{Part of the OpenFOAM software,
  \href{http://openfoam.org}{http://openfoam.org}}
implementation~\cite{ScaRooWhiDarRee} of the DSMC method. We modified
the dsmcFoam software implementations to output the stress and heat
flux inside the domain, in addition to the density, velocity and
temperature.

Observe that the DSMC method does not simulate the deterministic
interaction of realistic gas molecules; on the contrary, the molecular
collisions in the DSMC method are modeled similarly to those of the
artificial ``random gas'' in~\cite{Abr13} (with the exception that the
velocity of colliding molecules is also used in the collision
selection algorithm~\cite{Bird}). Because of this similarity between
the DSMC collisions and the random gas formulation in~\cite{Abr13},
the empirically determined mass diffusivity coefficient $D_\alpha$
(which we found suitable for the simulations studied below) is likely
underestimated versus a realistic gas with deterministic molecular
interactions.

For the dsmcFoam computations, we implemented a two-dimensional one
micrometer wide and four micrometers long channel, partitioned into
uniform rectangular cells, with 100 cells across the channel and 200
cells along the channel (such that the size of each cell is
20$\times$10 nanometers). The coordinate system was oriented so that
the $x$-axis was aligned with the direction of the channel (horizontal
direction in the figures below), while the $y$-axis pointed across the
channel (vertical direction in the figures below).

At the entrance of the channel, we specified the
following thermodynamic parameters for the DSMC simulation:
\begin{itemize}
\item The number density and temperature of the incoming gas flow was
  set to 10$^{25}$ molecules per cubic meter and 288.15 K (15$^\circ$
  C), respectively;
\item The velocity of the incoming gas flow was distributed
  parabolically across the channel, set to 20 meters per second at the
  walls, and to 100 meters per second in the middle of the channel.
\end{itemize}
These parameters were used for the simulation of both argon and
nitrogen. The resulting pressure was roughly one-half of that at sea
level at the entrance of the channel, and one-third of that at the
exit. For these conditions, the mean free path of a gas molecule was
roughly 150 nanometers long on average, so that the two Knudsen
boundary layers (one near each of the two walls) together occupied
about one-third of the total width of the channel. To obtain the
averaged macroscopic gas flow parameters, we ran the dsmcFoam
simulation for $3\cdot 10^{-5}$ seconds.

Due to the manifestation of computational artifacts in the dsmcFoam
simulations immediately adjacent to the entrance and exit of the
channel, for the subsequent gas dynamics simulations of the flow we
discarded the first and last 0.5 micrometers of the channel (so that
the benchmark DSMC flow would be ``clean'' throughout the remainder of
the channel). For this reason, the computational domain of the channel
for the gas dynamics simulations below is three micrometers long
(instead of four), and starts and ends at 0.5 and 3.5 micrometers,
respectively.

\begin{figure}
\includegraphics[width=0.6\textwidth]{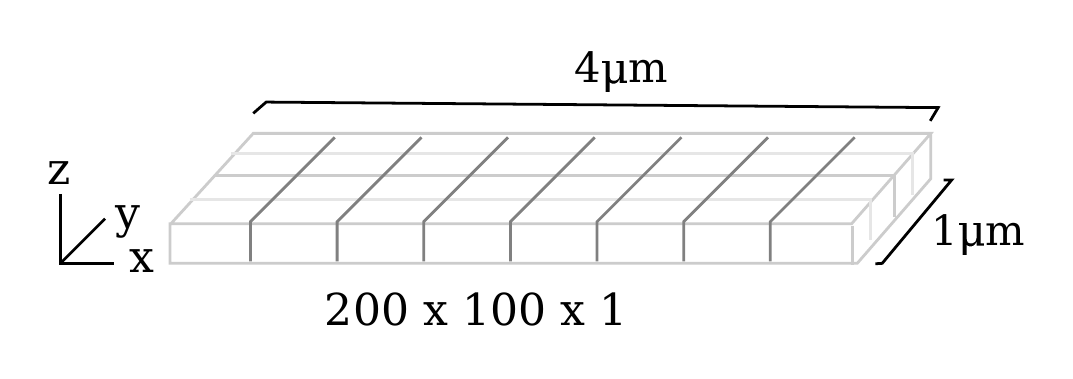}
\includegraphics[width=0.6\textwidth]{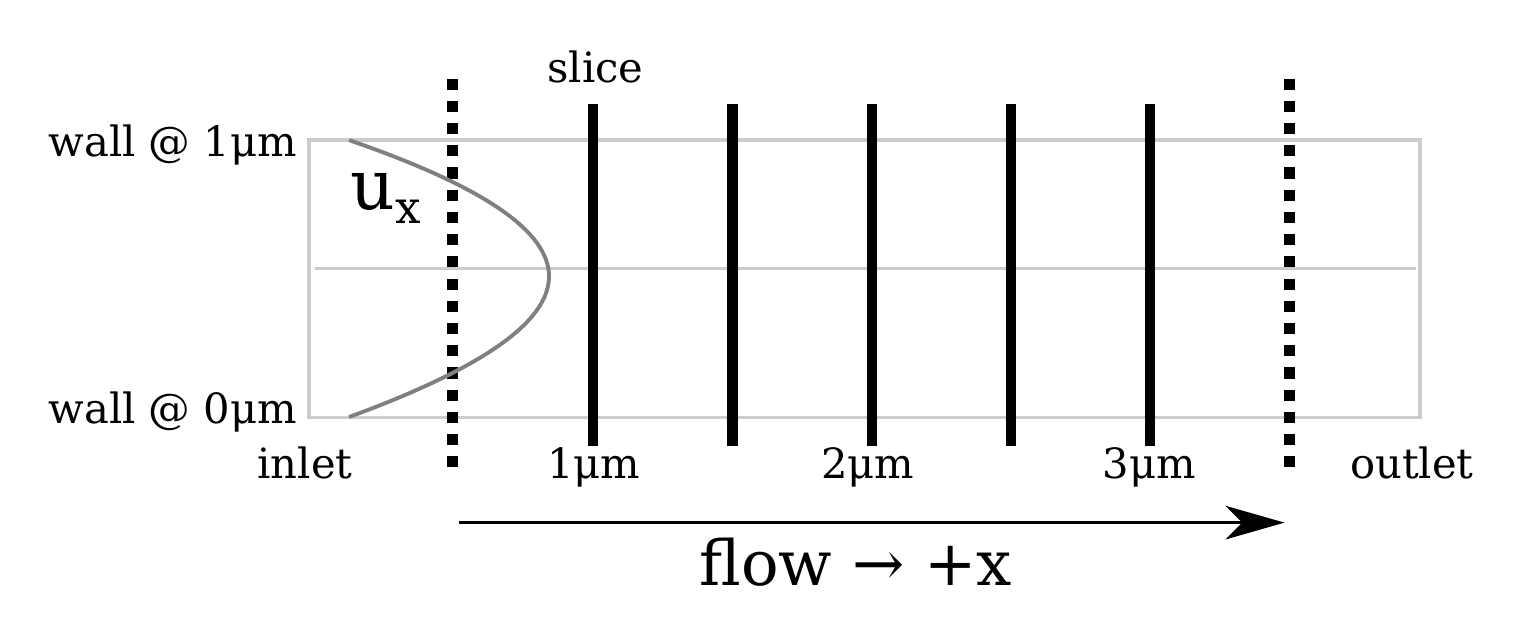}
\caption{A schematic representation of the finite volume mesh of the
  channel and the DSMC set-up for the Poiseuille flow.}
\label{fig:schematic}
\end{figure}

For a schematic representation of the finite volume mesh of the
channel and the DSMC computational set-up, see
Figure~\ref{fig:schematic}.

\subsection{The implementation of the continuum gas dynamics methods}

We implemented the Navier-Stokes equations (both conventional and
diffusive) and the diffusive regularized Grad equations in the
OpenFOAM finite volume framework~\cite{WelTabJasFur}. Due to this
reason, for the Navier-Stokes and Grad simulations we imported the
same finite volume mesh that was used for the dsmcFoam computations
described above. For the diffusive Navier-Stokes and regularized Grad
equations, we used the following set of boundary conditions:
\begin{itemize}
\item Velocity, temperature, stress and heat flux (the latter two for
  the Grad equations) -- the Dirichlet boundary conditions at all
  boundaries (entrance, exit and the walls), which were imported directly
  from the dsmcFoam output;
\item Pressure -- the Dirichlet boundary conditions at the exit and
  entrance of the channel (imported from the dsmcFoam output), and the
  Neumann (zero normal gradient) boundary conditions at the walls.
\end{itemize}
For the conventional Navier-Stokes equations, the set of boundary
conditions above is overdetermined, and thus we had to resort to the
following ``relaxed'' set of boundary conditions:
\begin{itemize}
\item Velocity -- the Dirichlet boundary conditions at the exit and
  the walls, the Neumann (zero normal gradient) boundary condition at
  the entrance;
\item Temperature -- the Dirichlet boundary conditions at all
  boundaries (entrance, exit and the walls);
\item Pressure -- the Dirichlet boundary condition at the entrance,
  the Neumann (zero normal gradient) boundary condition at the exit
  and the walls.
\end{itemize}
The reason for this choice of the combination of the Dirichlet and
Neumann boundary conditions for the pressure and velocity of the
conventional Navier-Stokes equations at the entrance and exit is the
qualitative correspondence to the dsmcFoam data; clearly, the pressure
gradient at the entrance and the velocity gradient at the exit are
nonzeros (see the figures below), and thus we chose to set the
Dirichlet boundary conditions for the pressure at the entrance and the
velocity at the exit. The corresponding zero-gradient Neumann boundary
conditions were then specified at the remaining boundaries.

The following notations are used in all subsequent figures for
different types of equations:
\begin{itemize}
\item ``{\sf NS*}'' -- the conventional Navier-Stokes equations
  (that is, \eqref{eq:rho_u_E}, \eqref{eq:p_theta}, and \eqref{eq:S_q_NS} with
  $D_\alpha=0$) without the near-wall viscosity scaling
  in~\eqref{eq:viscosity_scaling};
\item ``{\sf NS}'' -- the conventional Navier-Stokes equations with
  the near-wall viscosity scaling in~\eqref{eq:viscosity_scaling};
\item ``{\sf dNS}'' -- the diffusive Navier-Stokes equations (that is,
  \eqref{eq:rho_u_E}, \eqref{eq:p_theta}, and \eqref{eq:S_q_NS} with
  $D_\alpha >0$) with the near-wall viscosity scaling
  in~\eqref{eq:viscosity_scaling};
\item ``{\sf dRG}'' -- the diffusive regularized Grad equations (that
  is, \eqref{eq:rho_u_E}, \eqref{eq:p_theta}, \eqref{eq:S_q},
  \eqref{eq:Reg_Grad} and \eqref{eq:Reg_Grad_2} with $D_\alpha >0$)
  with the near-wall viscosity scaling
  in~\eqref{eq:viscosity_scaling};
\item ``{\sf dsmc}'' -- the dsmcFoam simulation.
\end{itemize}

\subsection{Argon}

\begin{figure}
\includegraphics[width=\textwidth]{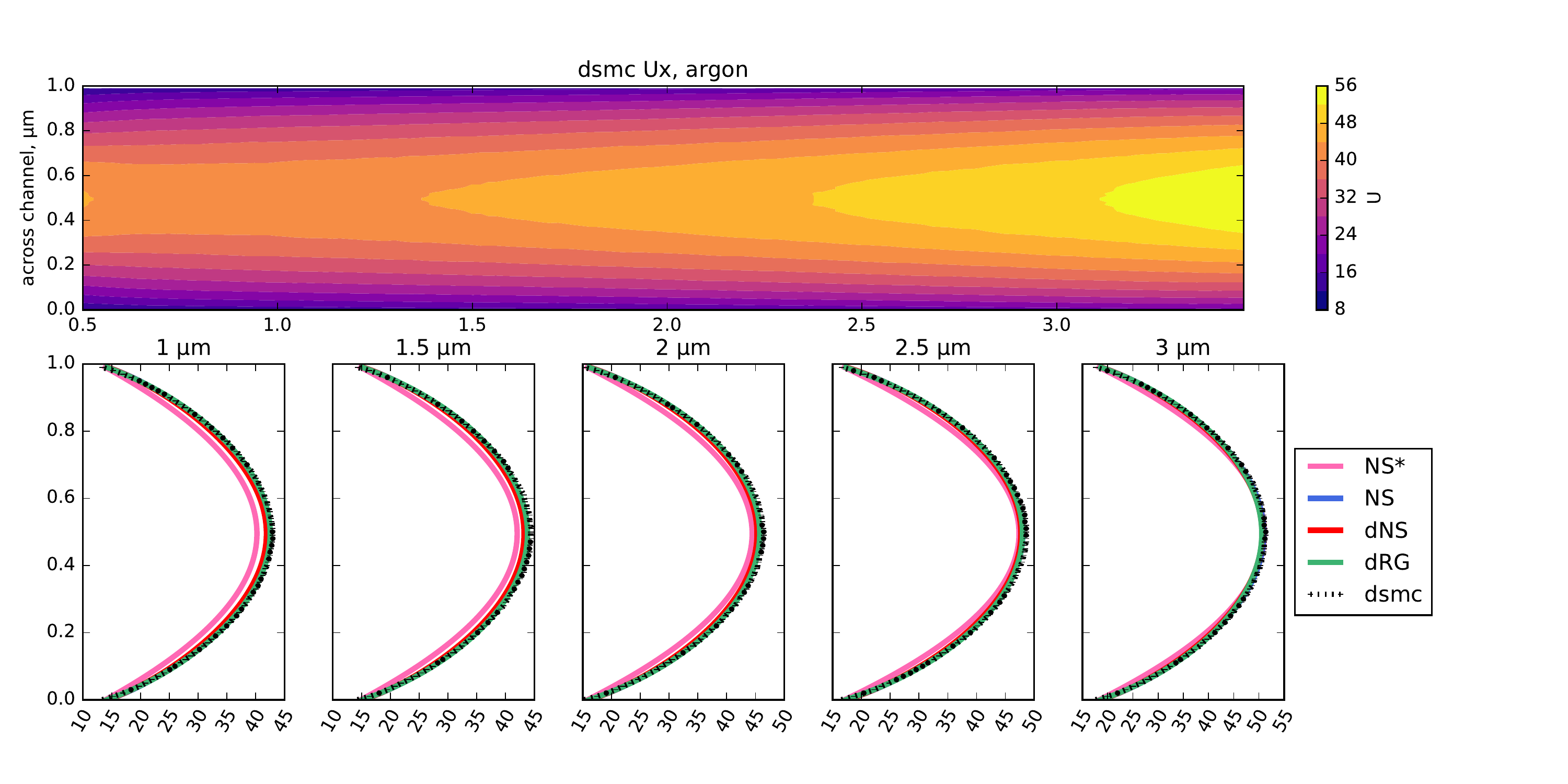}
\caption{Velocity of the Poiseuille flow of argon (m/s).}
\label{fig:argon_Ux}
\end{figure}

\begin{figure}
\includegraphics[width=\textwidth]{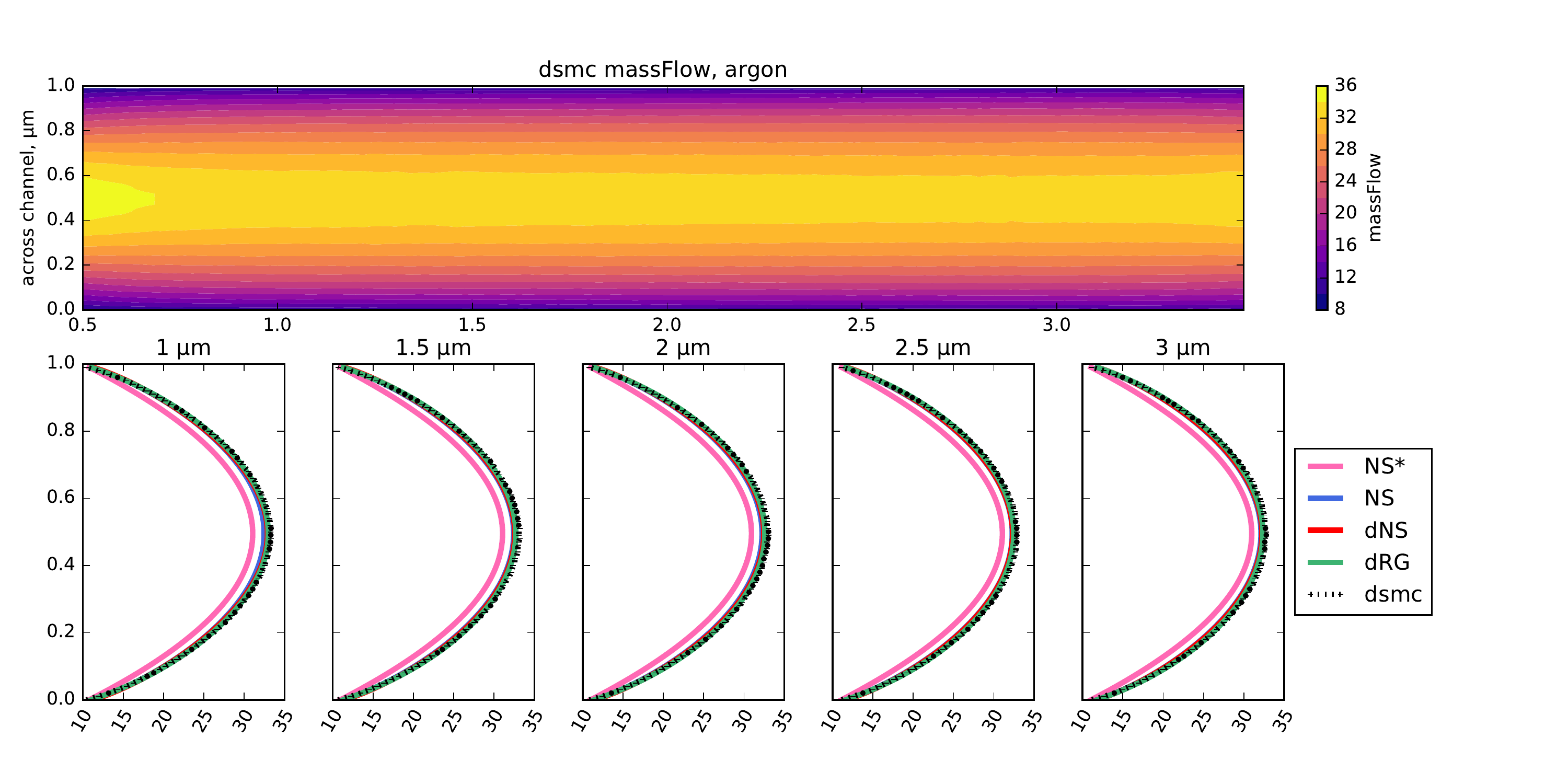}
\caption{Mass flow of the Poiseuille flow of argon (kg/(m$^2$ s)).}
\label{fig:argon_mf}
\end{figure}

\begin{figure}
\includegraphics[width=\textwidth]{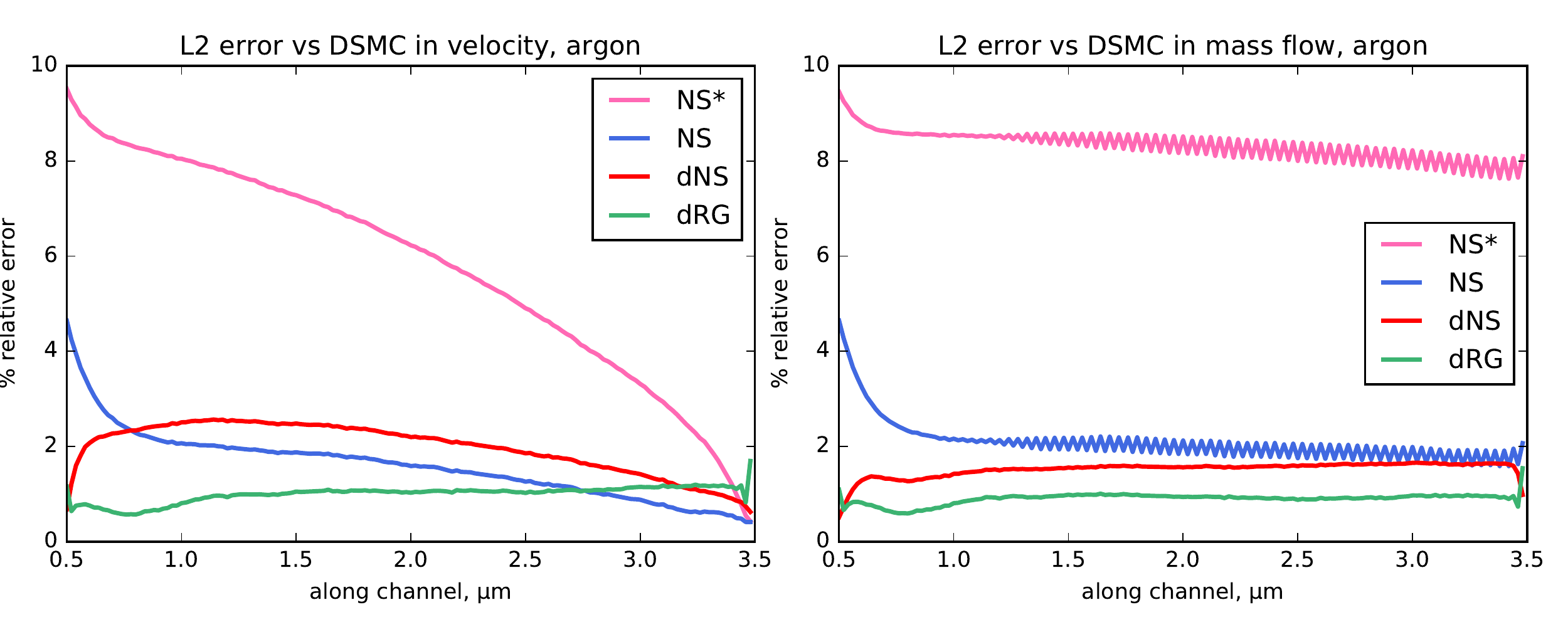}
\caption{Poiseuille flow of argon, errors in velocity and mass flow.}
\label{fig:argon_errors_Ux_mf}
\end{figure}

\begin{figure}
\includegraphics[width=\textwidth]{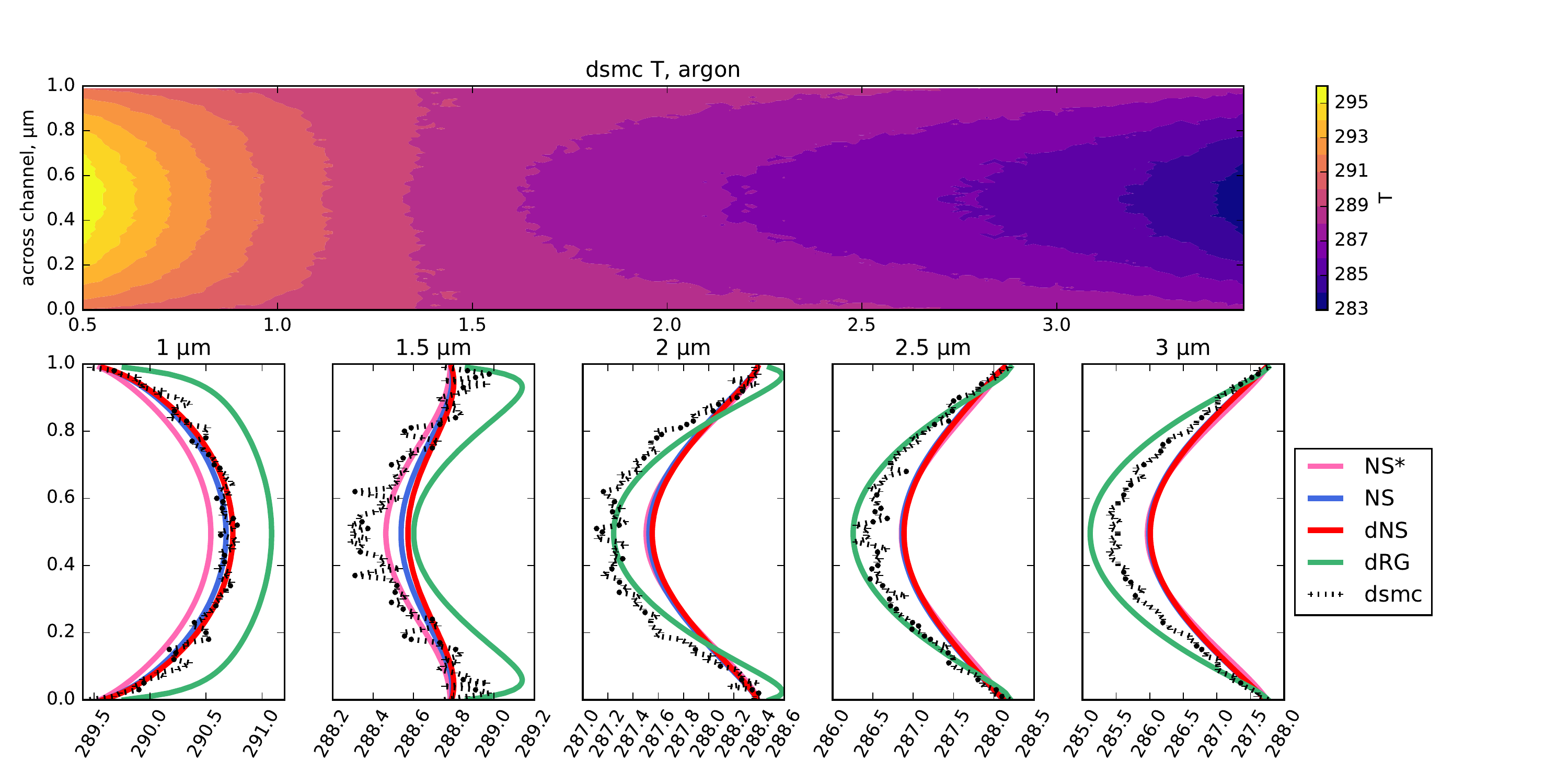}
\caption{Temperature of the Poiseuille flow of argon (K).}
\label{fig:argon_T}
\end{figure}

\begin{figure}
\includegraphics[width=\textwidth]{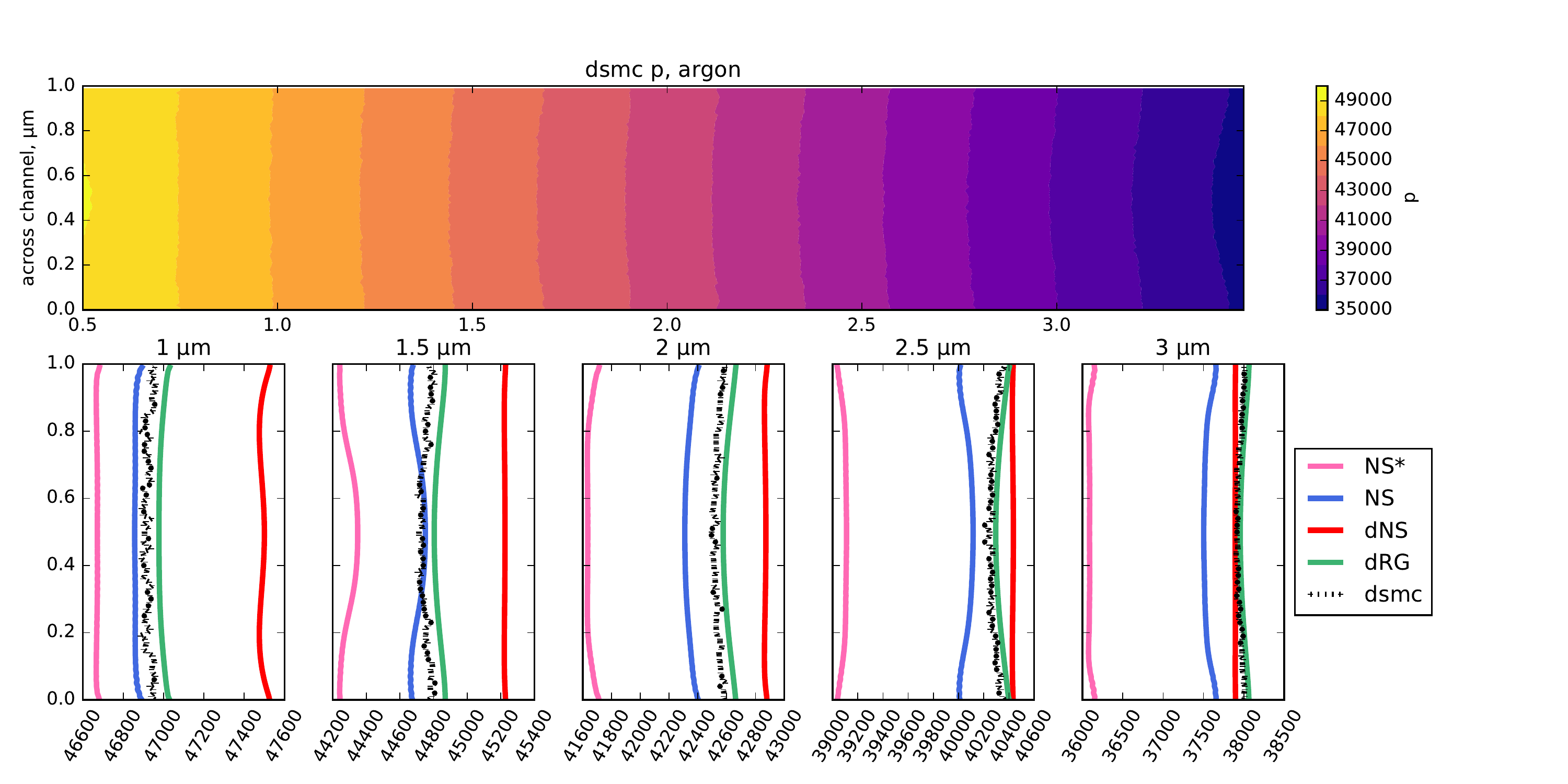}
\caption{Pressure of the Poiseuille flow of argon (kg/(m s$^2$)).}
\label{fig:argon_p}
\end{figure}

\begin{figure}
\includegraphics[width=\textwidth]{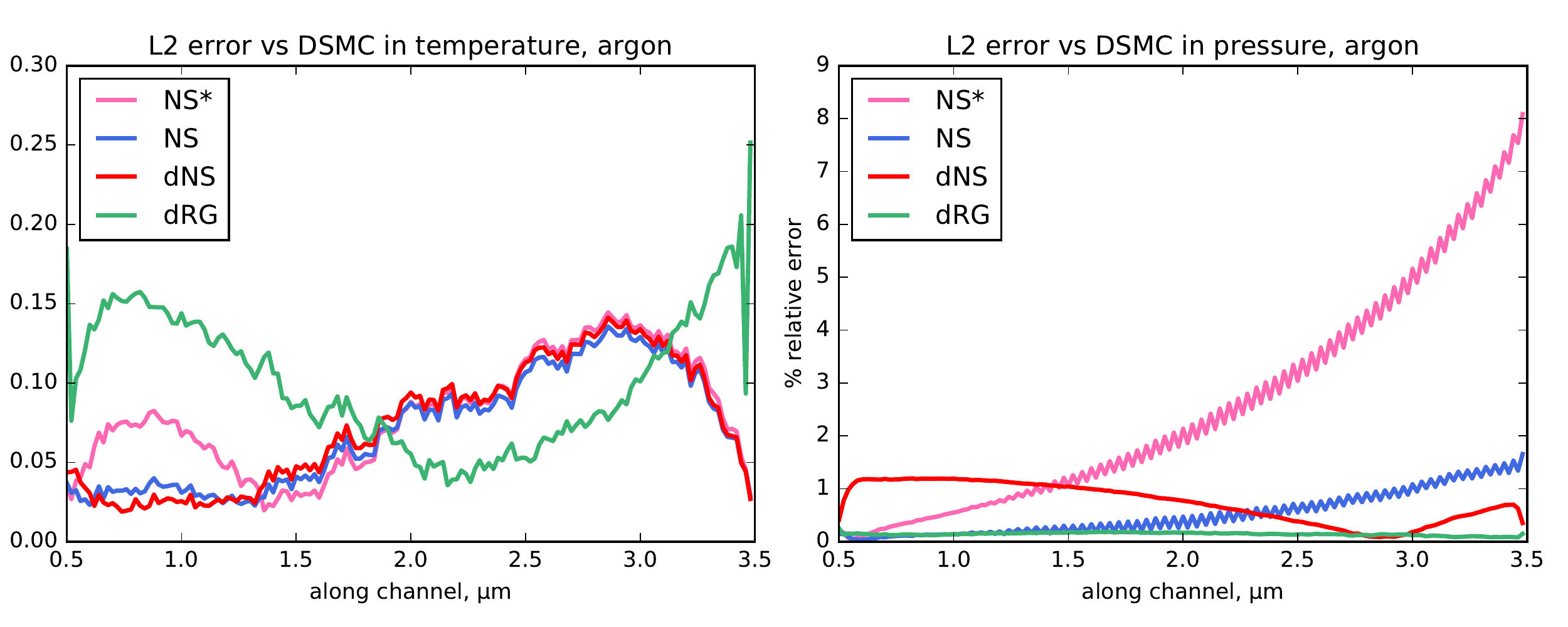}
\caption{Poiseuille flow of argon, errors in temperature and pressure.}
\label{fig:argon_errors_Tp}
\end{figure}

\begin{figure}
\includegraphics[width=\textwidth]{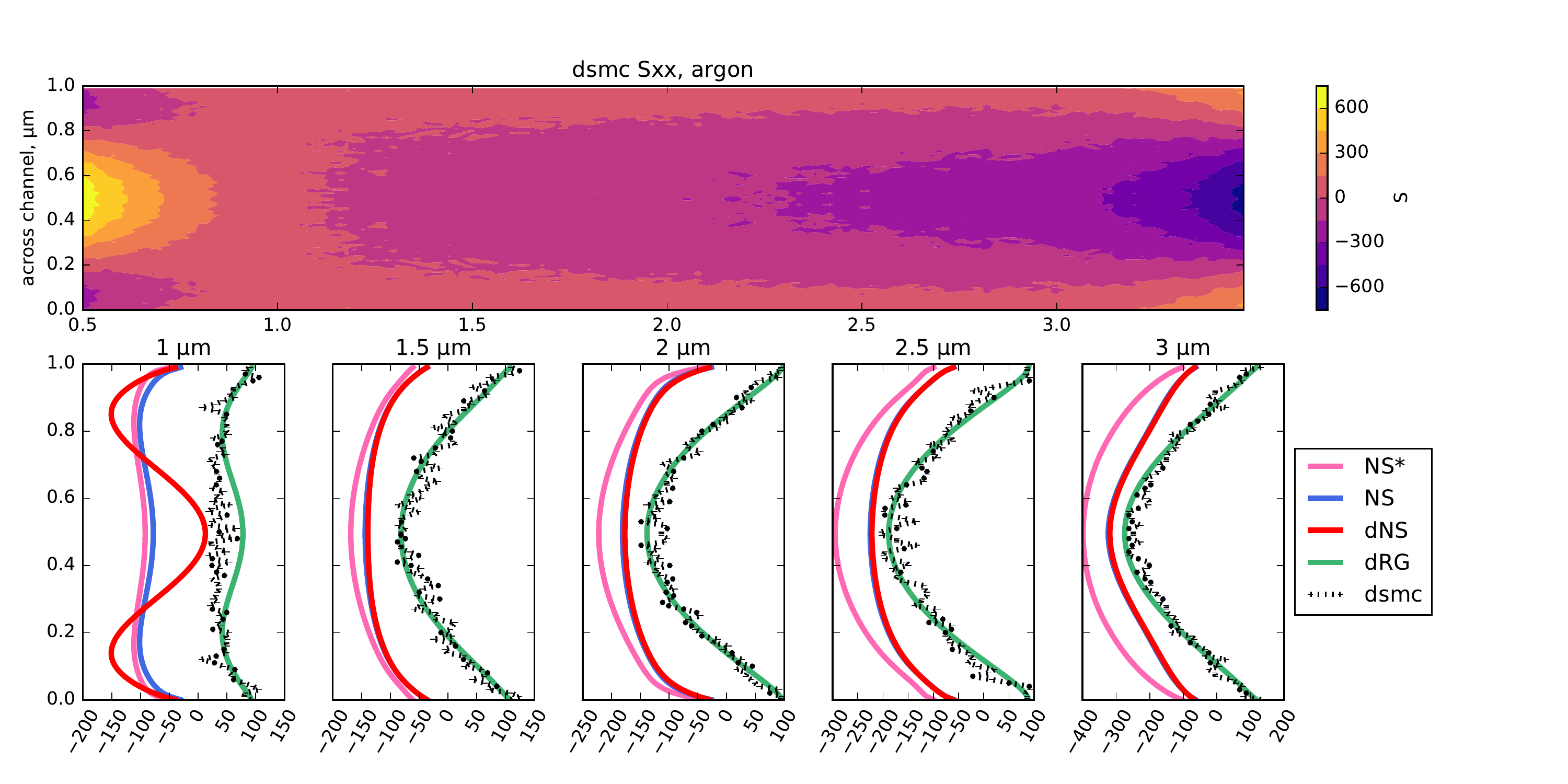}
\caption{Stress component $S_{xx}$ of the Poiseuille flow of
  argon (m$^2$/s$^2$).}
\label{fig:argon_Sxx}
\end{figure}

\begin{figure}
\includegraphics[width=\textwidth]{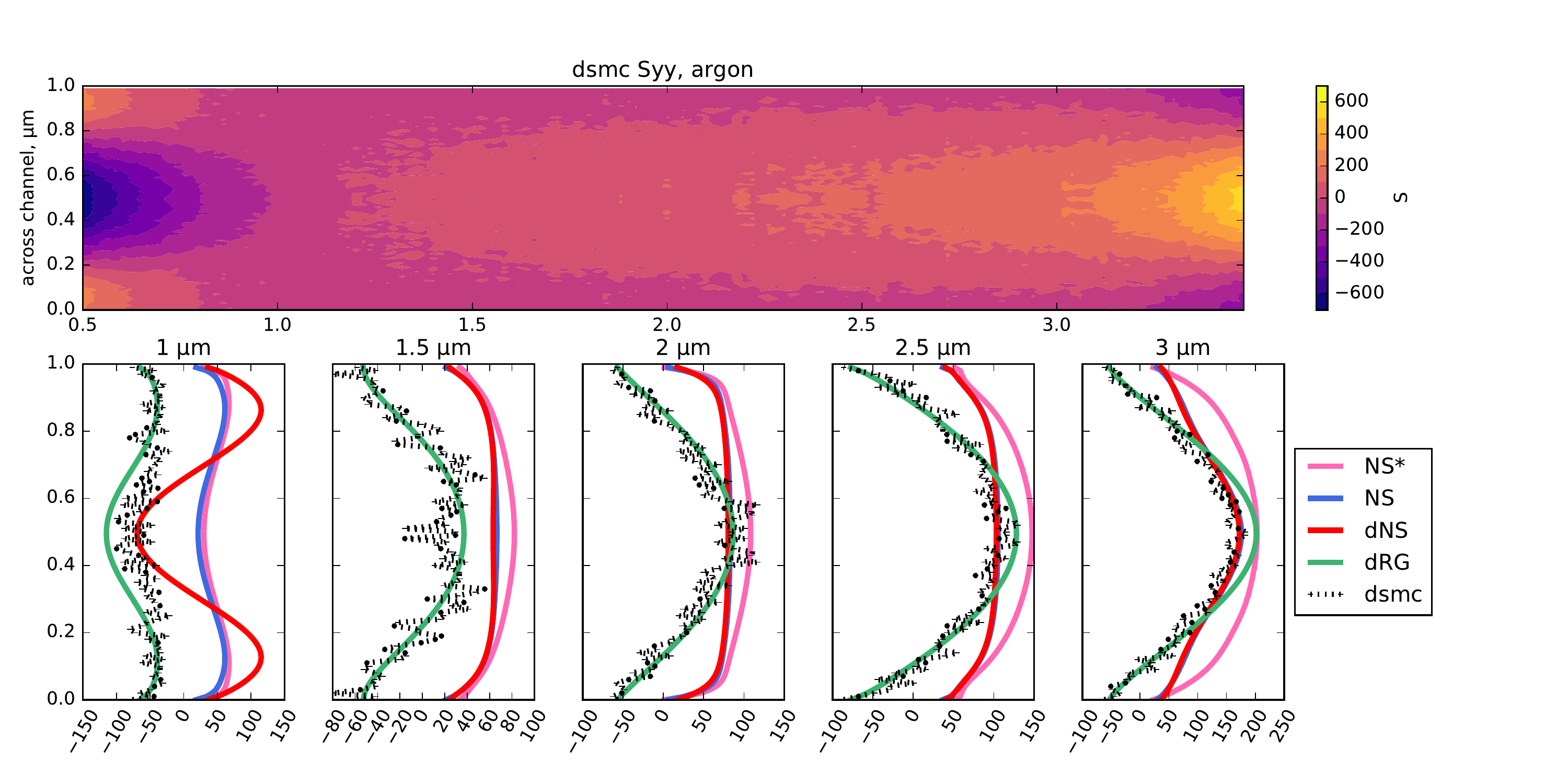}
\caption{Stress component $S_{yy}$ of the Poiseuille flow of
  argon (m$^2$/s$^2$).}
\label{fig:argon_Syy}
\end{figure}

\begin{figure}
\includegraphics[width=\textwidth]{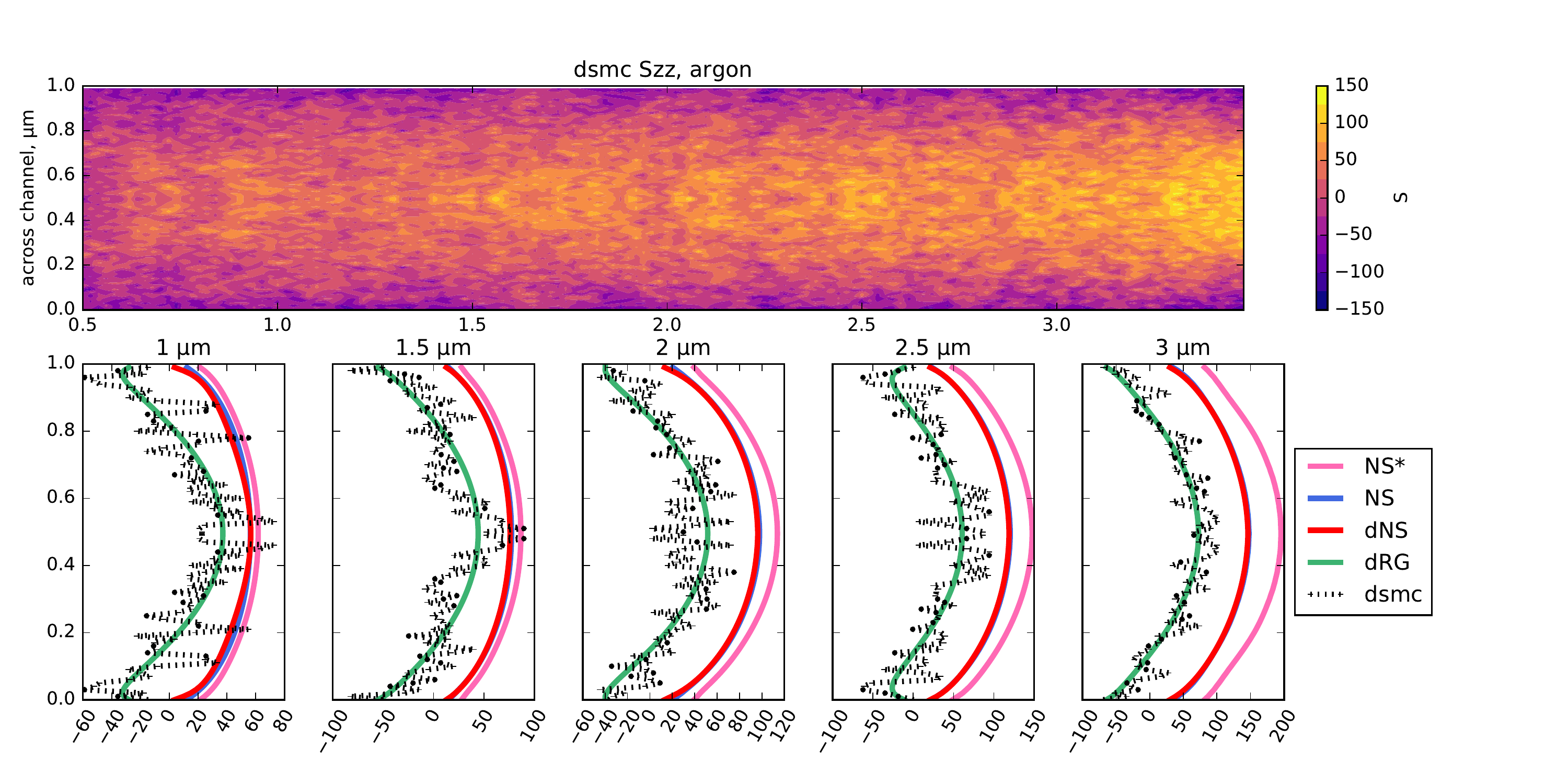}
\caption{Stress component $S_{zz}$ of the Poiseuille flow of
  argon (m$^2$/s$^2$).}
\label{fig:argon_Szz}
\end{figure}

\begin{figure}
\includegraphics[width=\textwidth]{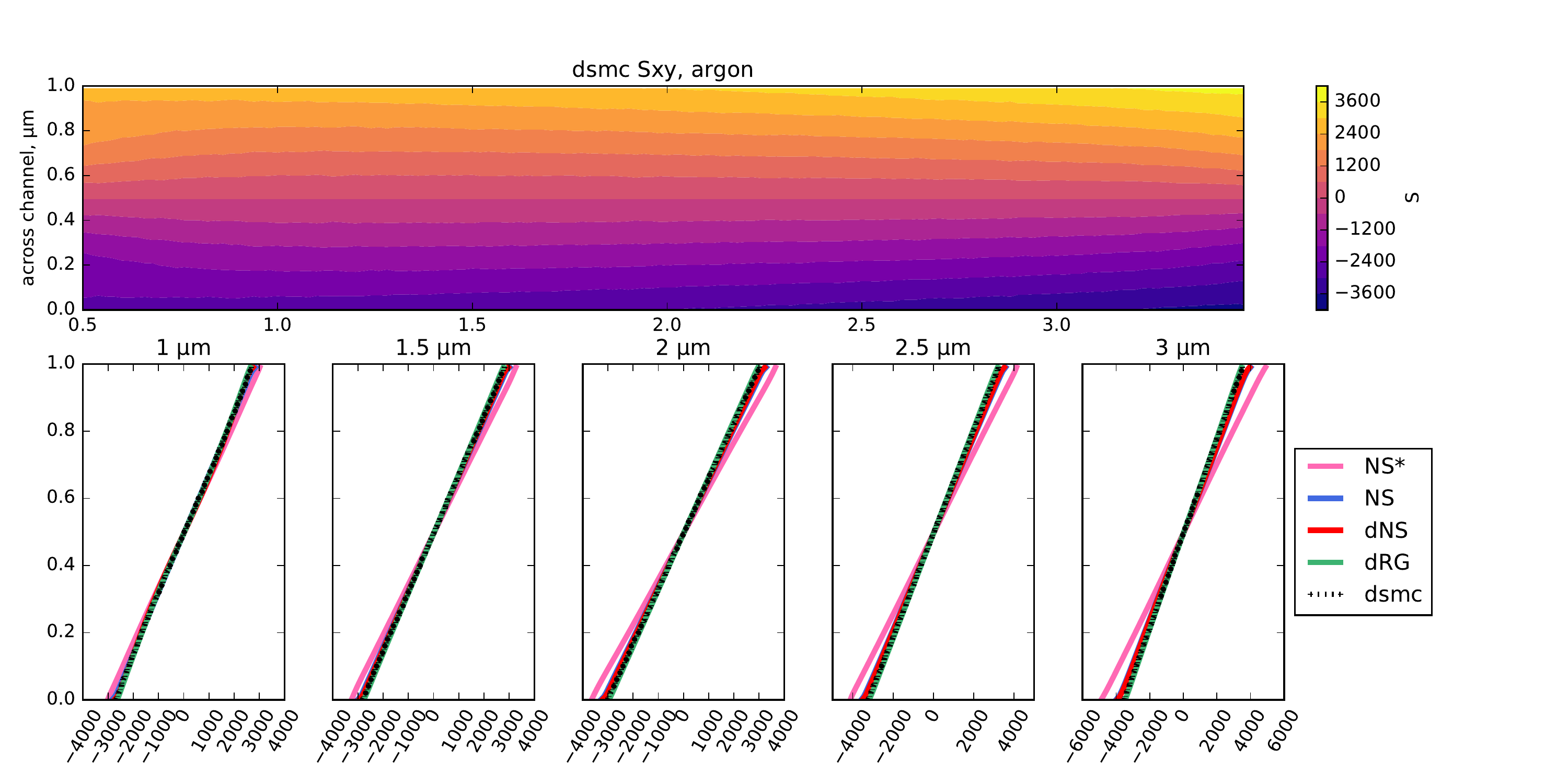}
\caption{Stress component $S_{xy}$ of the Poiseuille flow of
  argon (m$^2$/s$^2$).}
\label{fig:argon_Sxy}
\end{figure}

\begin{figure}
\includegraphics[width=\textwidth]{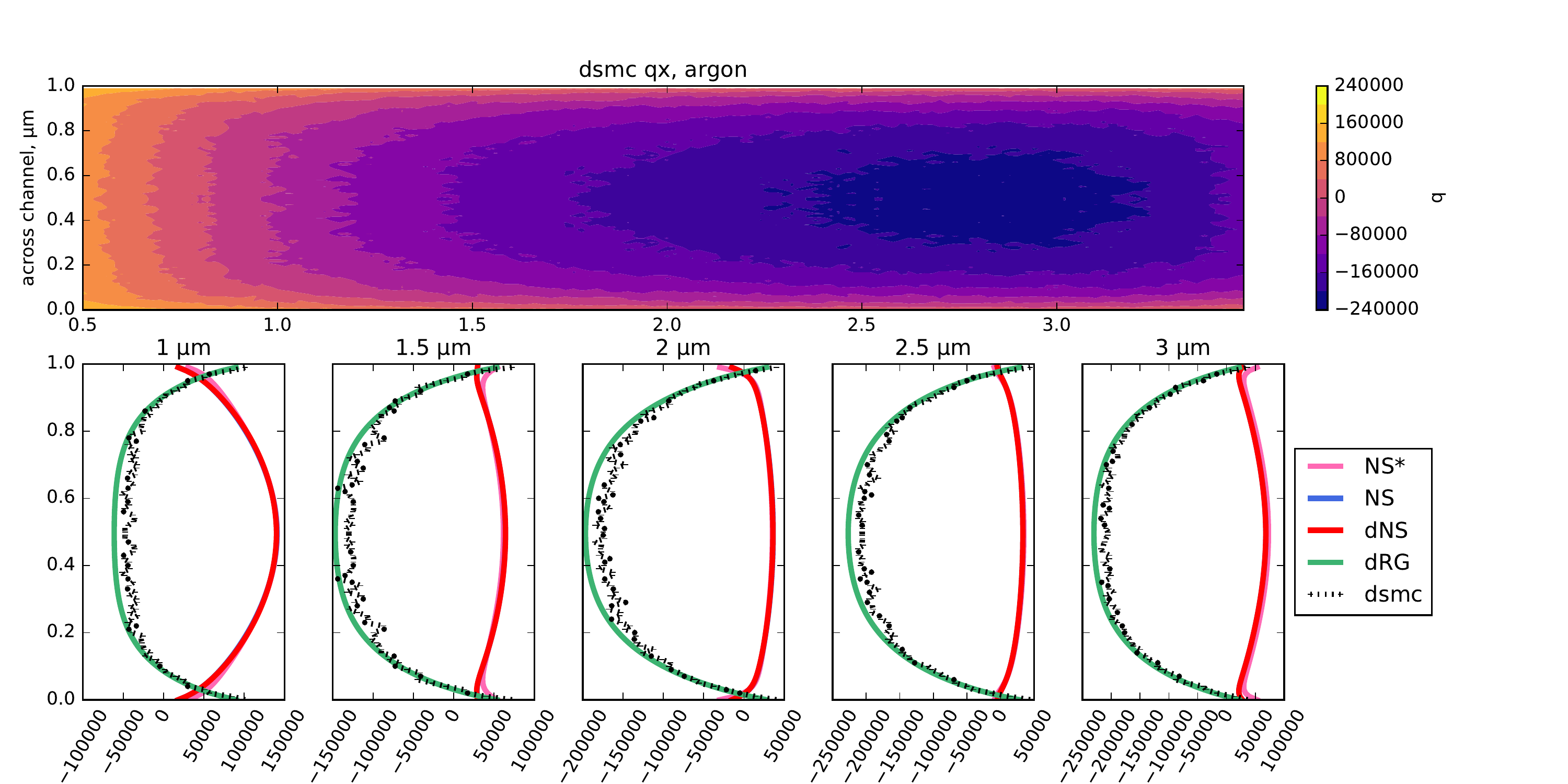}
\caption{Heat flux component $q_x$ of the Poiseuille flow of
  argon (m$^3$/s$^3$).}
\label{fig:argon_qx}
\end{figure}

\begin{figure}
\includegraphics[width=\textwidth]{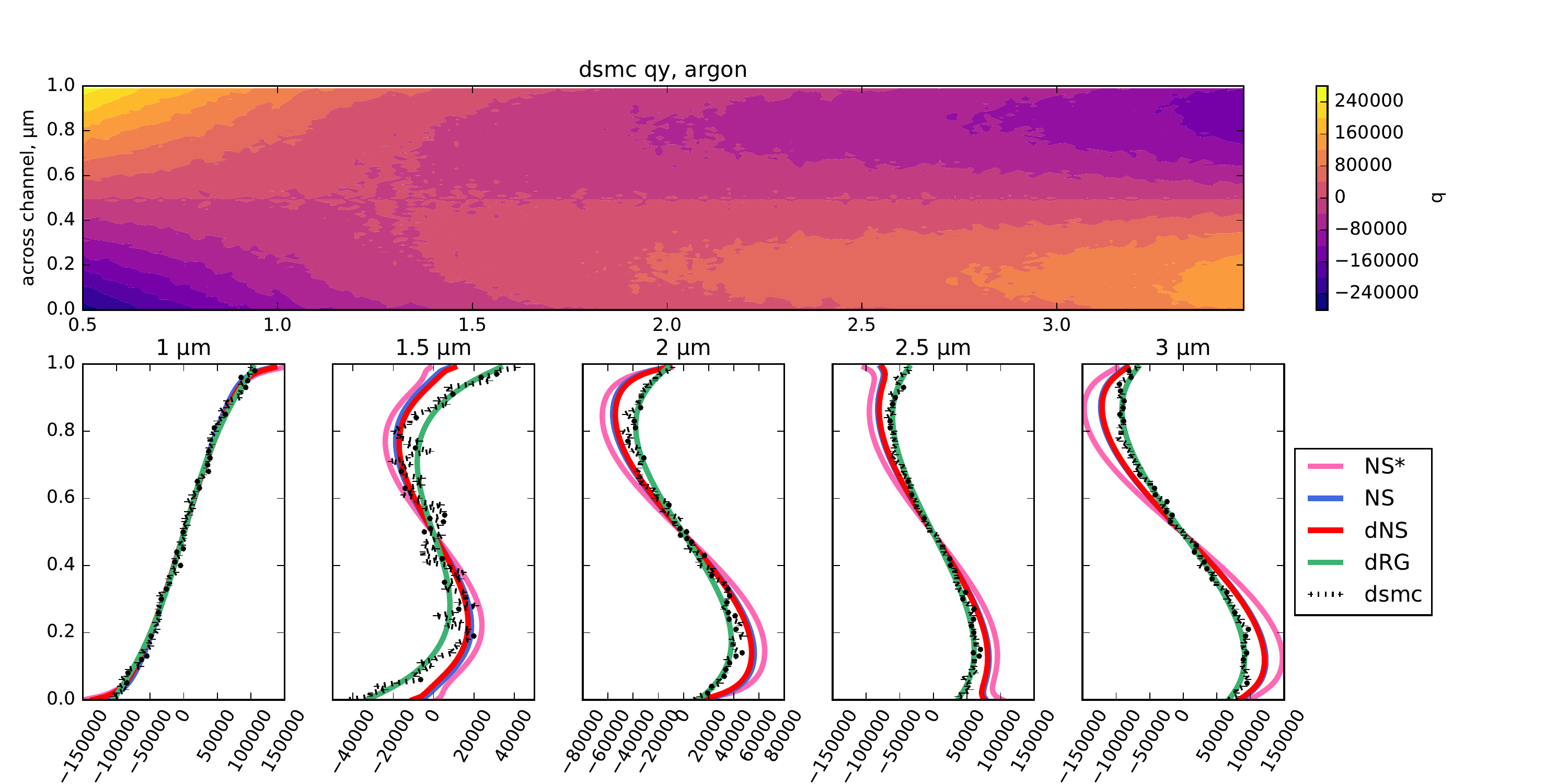}
\caption{Heat flux component $q_x$ of the Poiseuille flow of
  argon (m$^3$/s$^3$).}
\label{fig:argon_qy}
\end{figure}

Argon is a monatomic gas, which is often used in computational gas
dynamics benchmarks due to its nearly ideal properties. We used the
following computational parameters for argon:
\begin{itemize}
\item The adiabatic constant was set to $\gamma=5/3$,
\item The Prandtl number was set to $\Pran=2/3$,
\item The higher-order Prandtl numbers for the regularized Grad
  equations were set to $\Pran_{\widetilde{\!\BM Q}}=3/2$,
  $\Pran_{\widetilde R}=2/3$ and $\Pran_{\widetilde{\!\BM R}}=7/6$
  (for more details, see \cite{Stru,StruTor,TorStru}),
\item The molar mass $M$ was set to $3.995\cdot 10^{-2}$ kg/mol,
\item The viscosity $\mu$ and empirically scaled mass diffusivity
  $D_\alpha$ were proportional to the square root of the temperature:
\begin{equation}
\label{eq:mu_D}
\mu=\mu^*\sqrt{\frac{M\theta}{RT^*}},\qquad
D_\alpha=D_\alpha^*\sqrt{\frac{M\theta}{RT^*}},
\end{equation}
with the reference temperature set to $T^*=288.15$ K, whereas the
reference viscosity and empirical scaled mass diffusivity constants
were set to $\mu^*=2.2\cdot 10^{-5}$ kg/(m sec), and
$D_\alpha^*=10^{-6}$ kg/(m sec), respectively. The notation $R$ above
refers to the universal gas constant, $R=8.314$ kg m$^2$/(mol K
sec$^2$).
\end{itemize}
In Figures \ref{fig:argon_Ux}--\ref{fig:argon_errors_Tp} we show the
velocity, mass flow (the product of the density with velocity),
temperature and pressure for the continuum gas dynamics closures and
compare them against the DSMC computation. For each variable, we show
five evenly spaced profiles across the channel for all studied
closures, as well as the corresponding relative errors in
across-channel profiles as functions of the distance along the
channel. Observe that the conventional Navier-Stokes closure without
the viscosity scaling in~\eqref{eq:viscosity_scaling} is consistently
the least accurate continuum gas dynamics approximation among all
tested -- its relative error against the DSMC computation reaches
9-10\% in velocity, mass flow and pressure. On the other hand, the
diffusive regularized Grad closure with the viscosity scaling
in~\eqref{eq:viscosity_scaling} is consistently the most accurate
continuum gas dynamics closure -- its relative error against the DSMC
computation is about 1\% in velocity and mass flow, and about 0.2\% in
pressure. The temperature is approximated well by all continuum gas
dynamics closures -- the errors in temperature are about 0.2\%
irrespective of the closure (and is likely due to natural errors of
the finite volume approximation of differential operators in general).

\begin{figure}
\includegraphics[width=\textwidth]{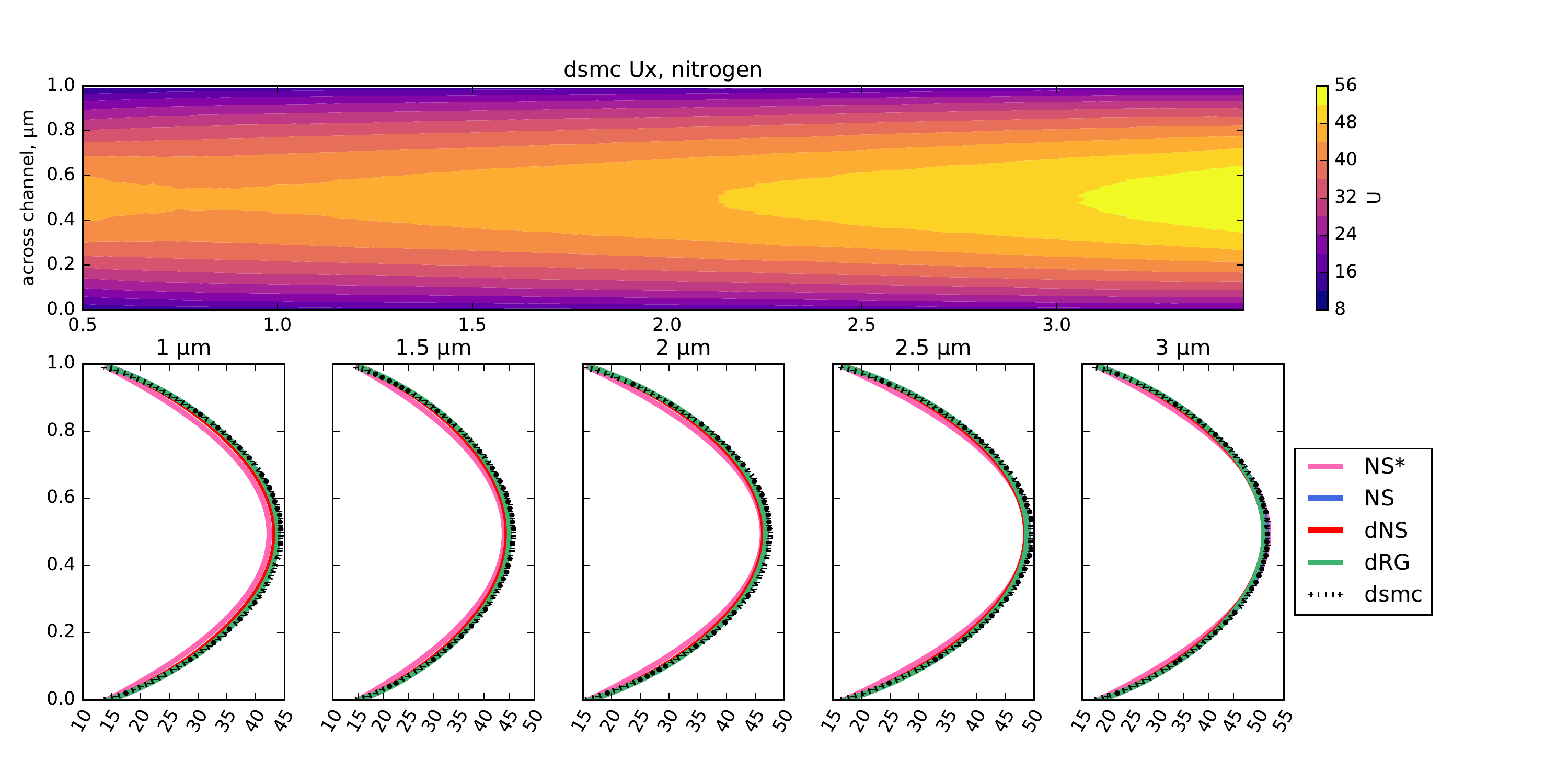}
\caption{Velocity of the Poiseuille flow of nitrogen (m/s).}
\label{fig:nitrogen_Ux}
\end{figure}

\begin{figure}
\includegraphics[width=\textwidth]{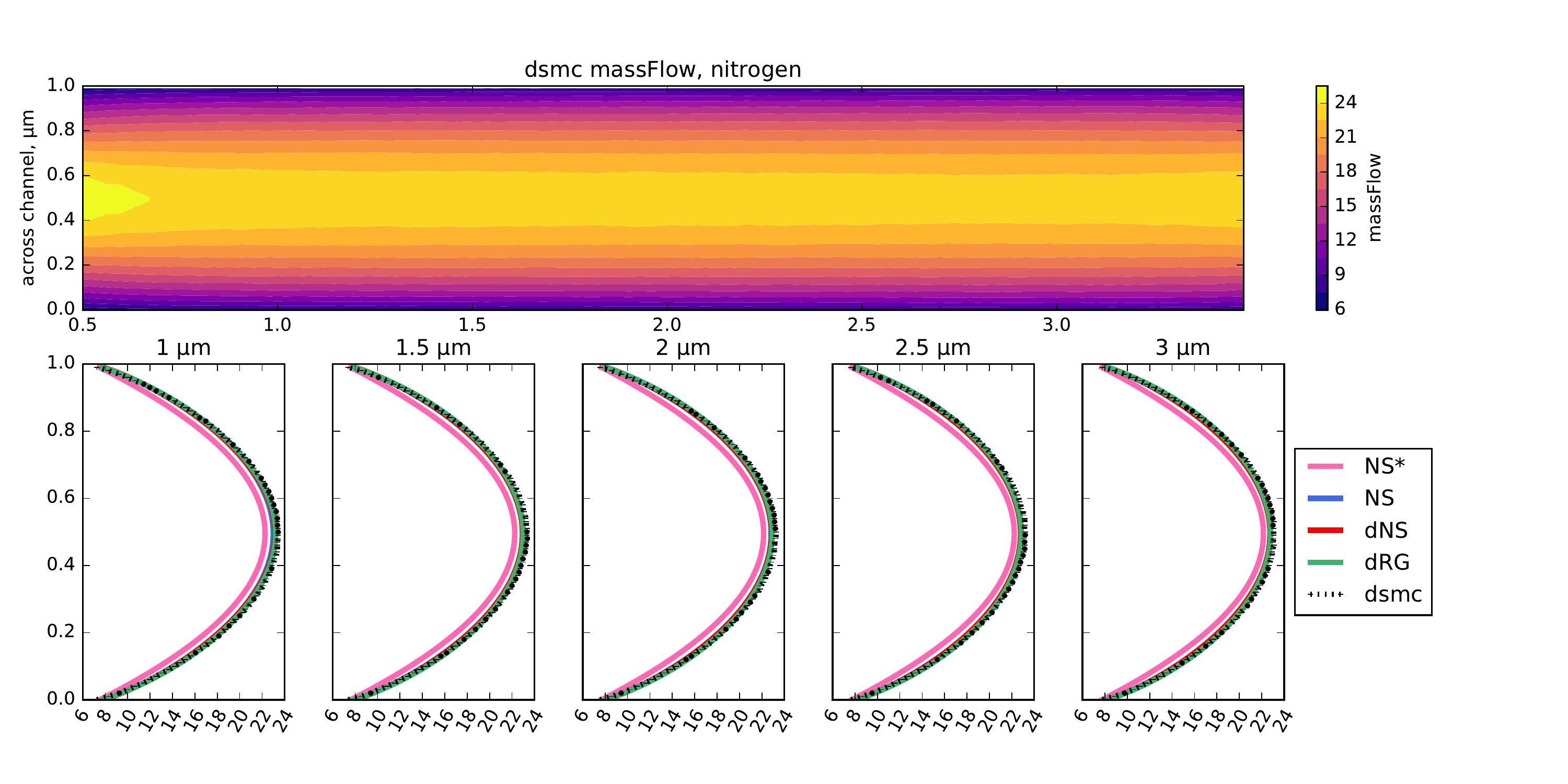}
\caption{Mass flow of the Poiseuille flow of nitrogen (kg/(m$^2$ s)).}
\label{fig:nitrogen_mf}
\end{figure}

\begin{figure}
\includegraphics[width=\textwidth]{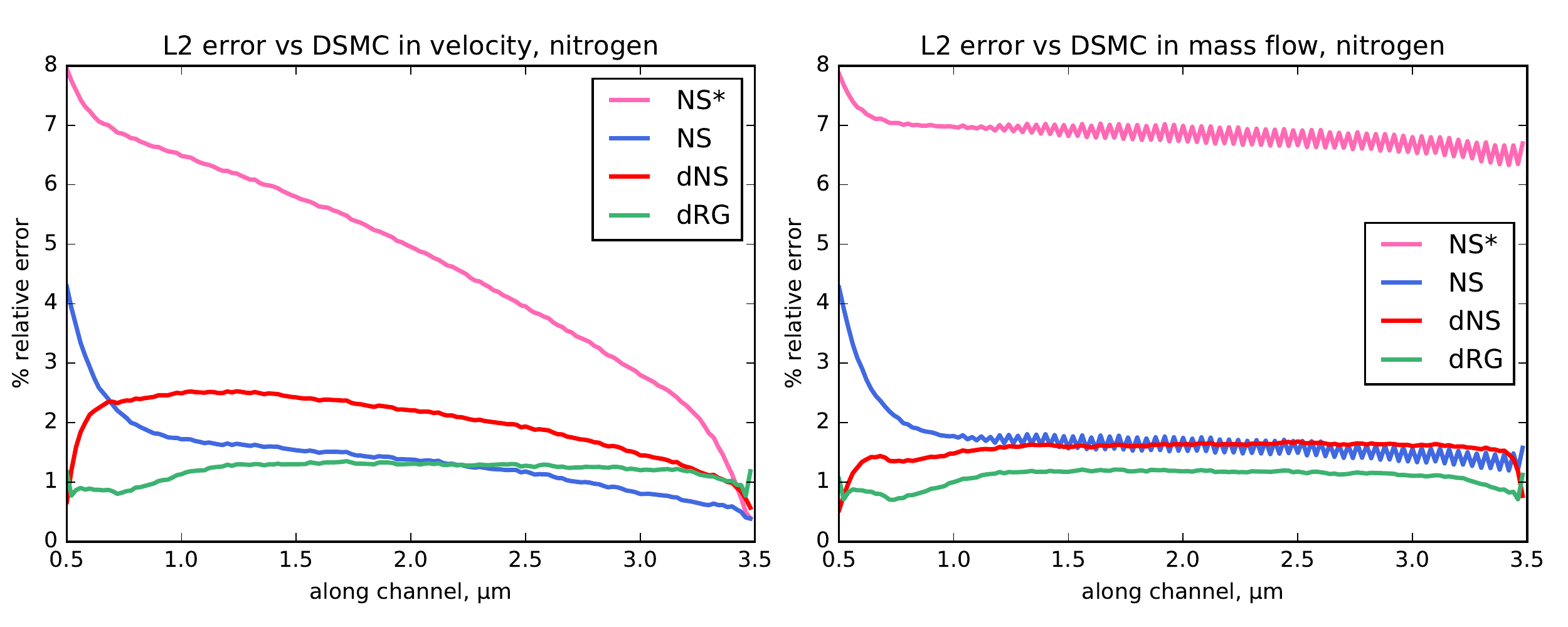}
\caption{Poiseuille flow of nitrogen, errors in velocity and mass flow.}
\label{fig:nitrogen_errors_Ux_mf}
\end{figure}

\begin{figure}
\includegraphics[width=\textwidth]{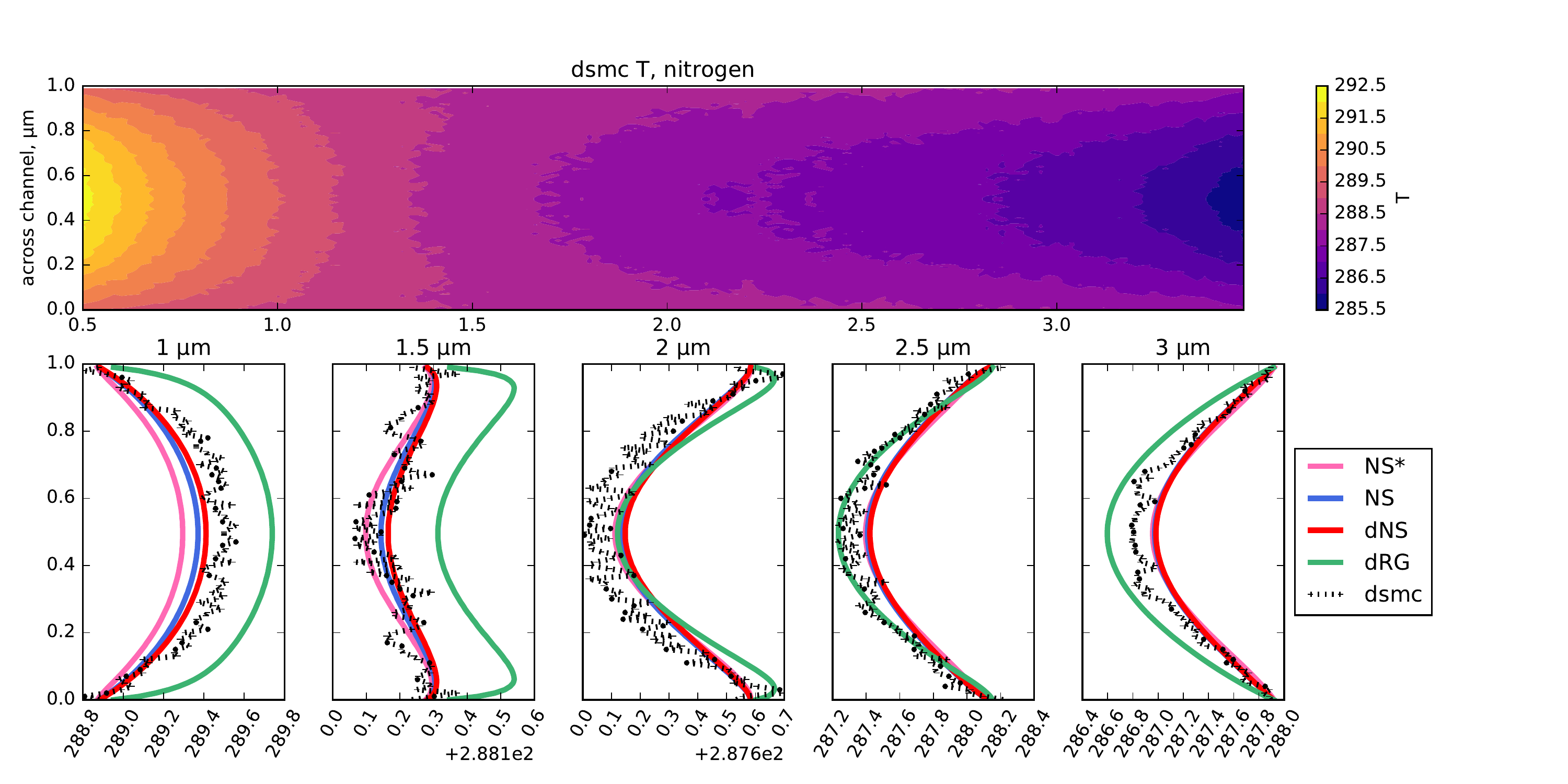}
\caption{Temperature of the Poiseuille flow of nitrogen (K).}
\label{fig:nitrogen_T}
\end{figure}

\begin{figure}
\includegraphics[width=\textwidth]{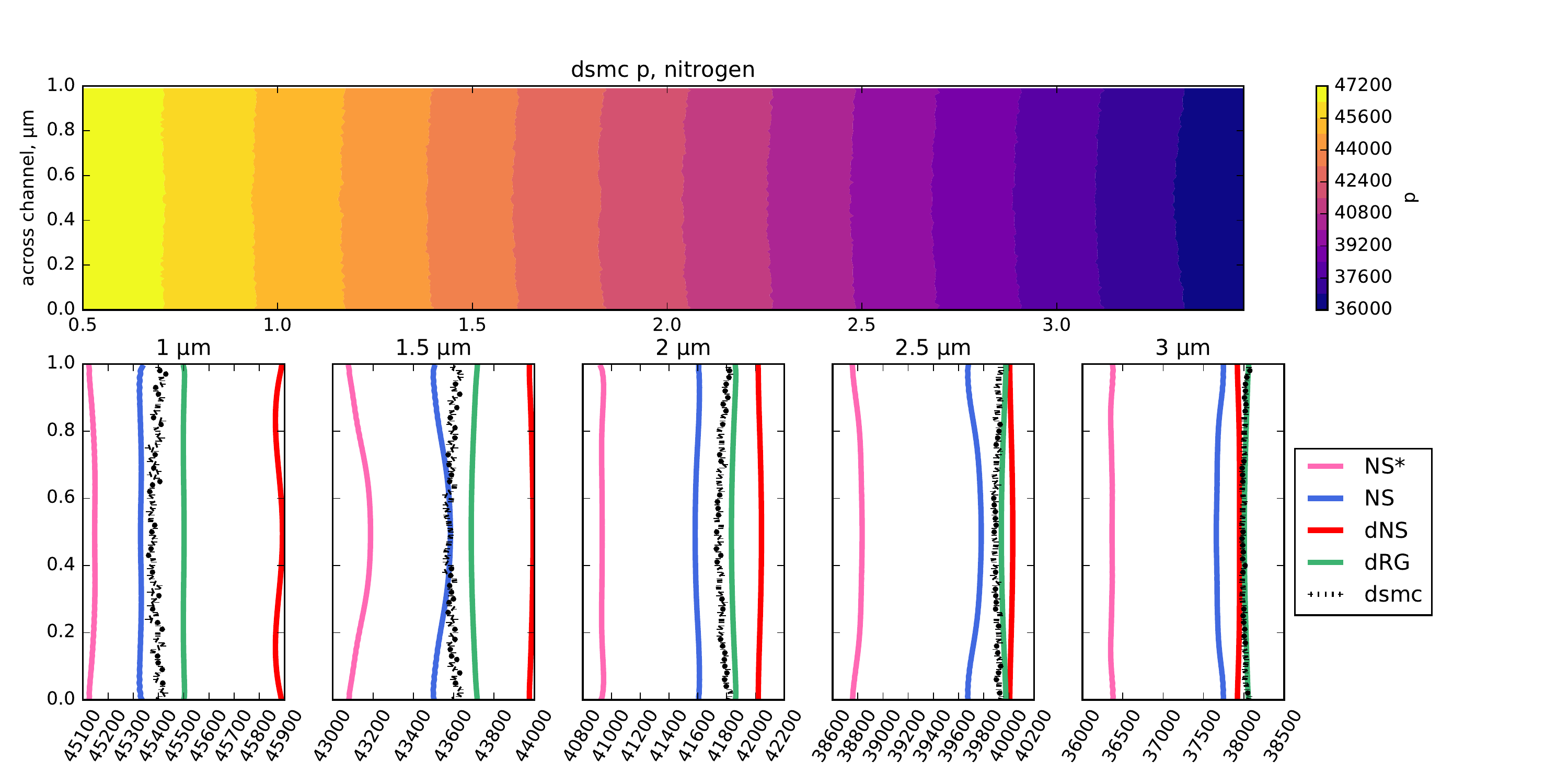}
\caption{Pressure of the Poiseuille flow of nitrogen (kg/(m s$^2$)).}
\label{fig:nitrogen_p}
\end{figure}

\begin{figure}
\includegraphics[width=\textwidth]{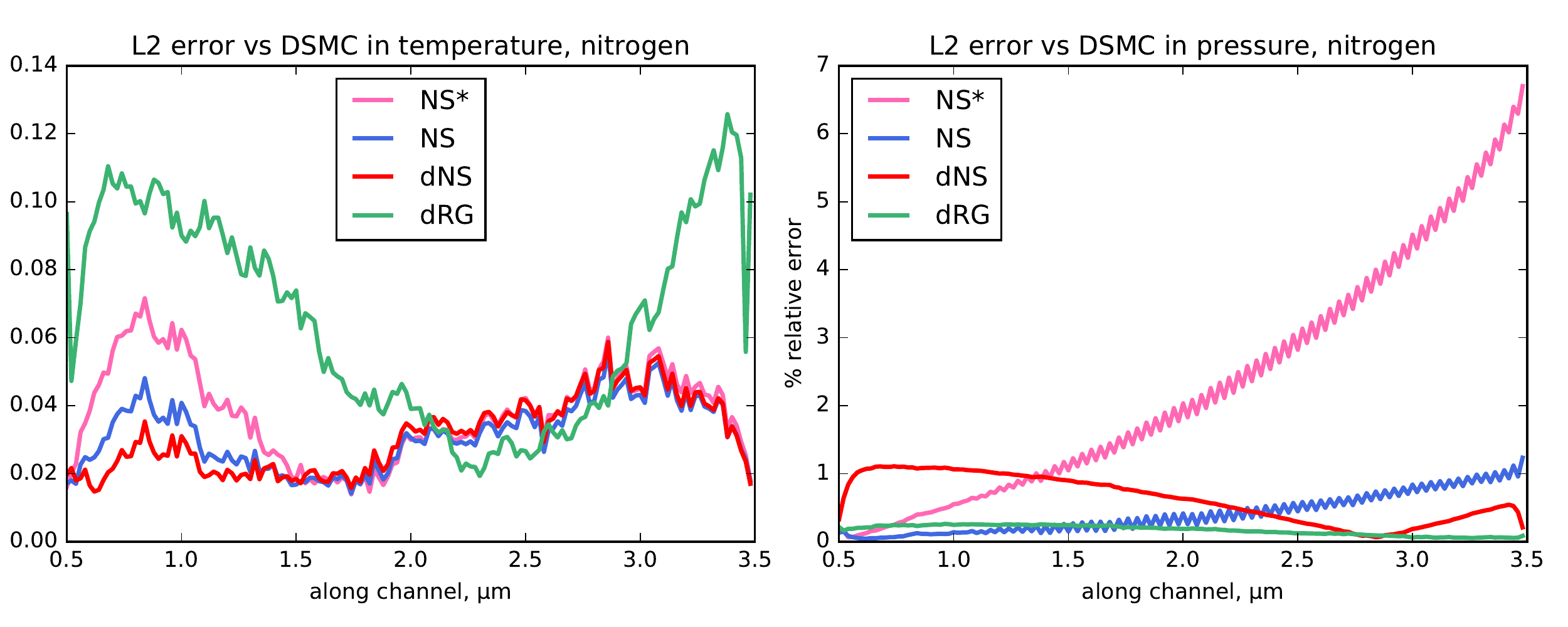}
\caption{Poiseuille flow of nitrogen, errors in temperature and pressure.}
\label{fig:nitrogen_errors_Tp}
\end{figure}

\begin{figure}
\includegraphics[width=\textwidth]{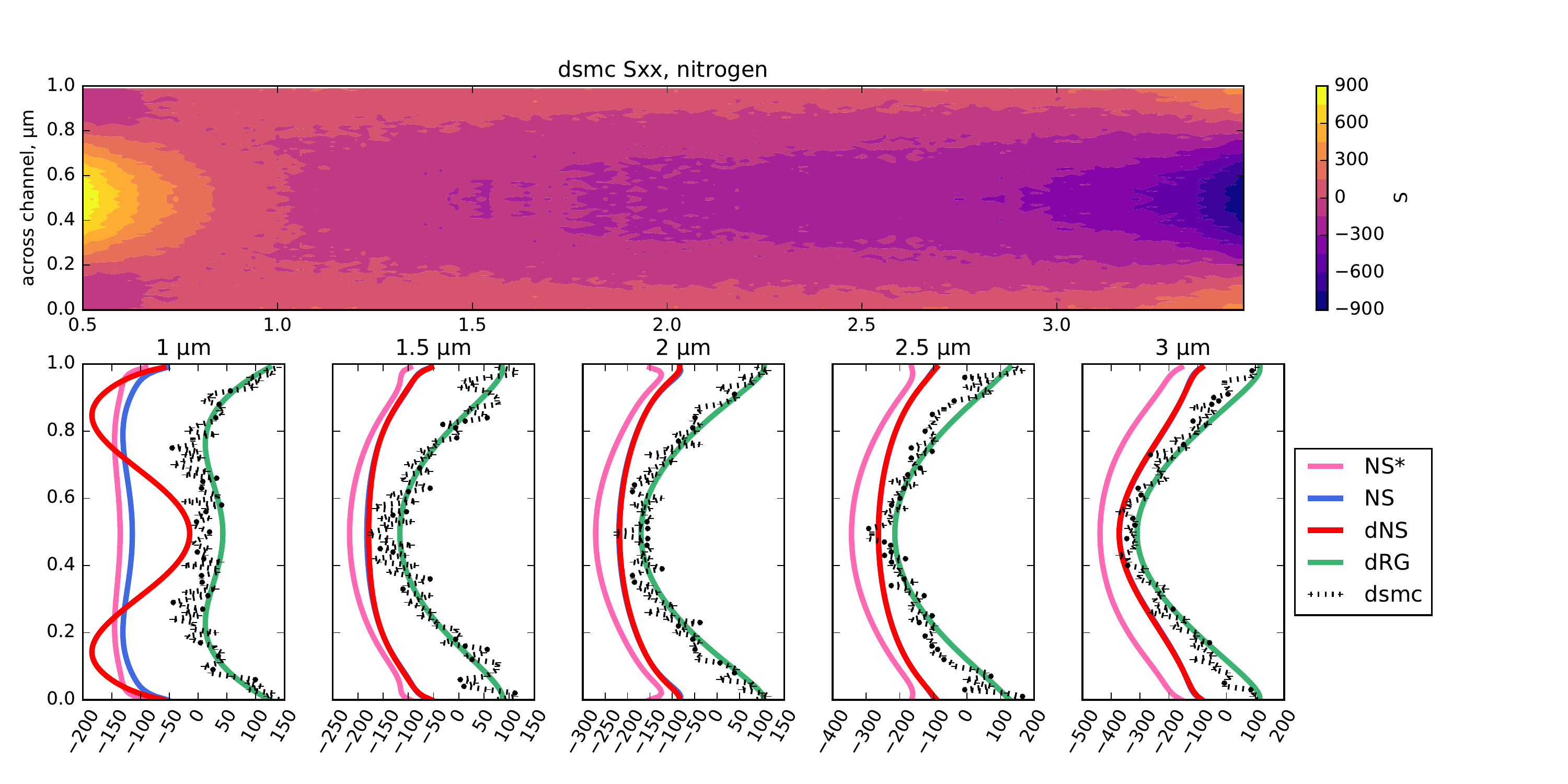}
\caption{Stress component $S_{xx}$ of the Poiseuille flow of
  nitrogen (m$^2$/s$^2$).}
\label{fig:nitrogen_Sxx}
\end{figure}

\begin{figure}
\includegraphics[width=\textwidth]{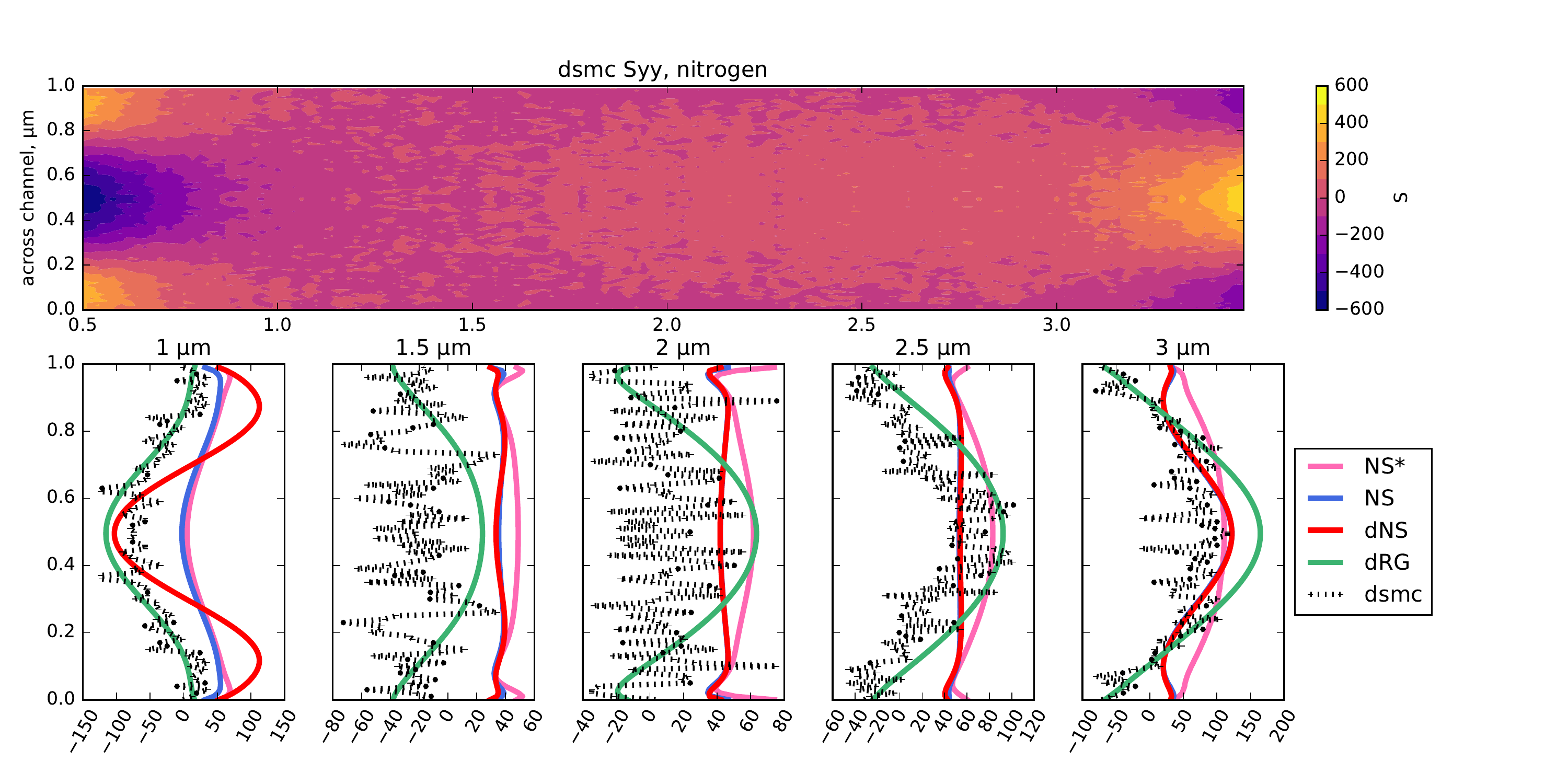}
\caption{Stress component $S_{yy}$ of the Poiseuille flow of
  nitrogen (m$^2$/s$^2$).}
\label{fig:nitrogen_Syy}
\end{figure}

\begin{figure}
\includegraphics[width=\textwidth]{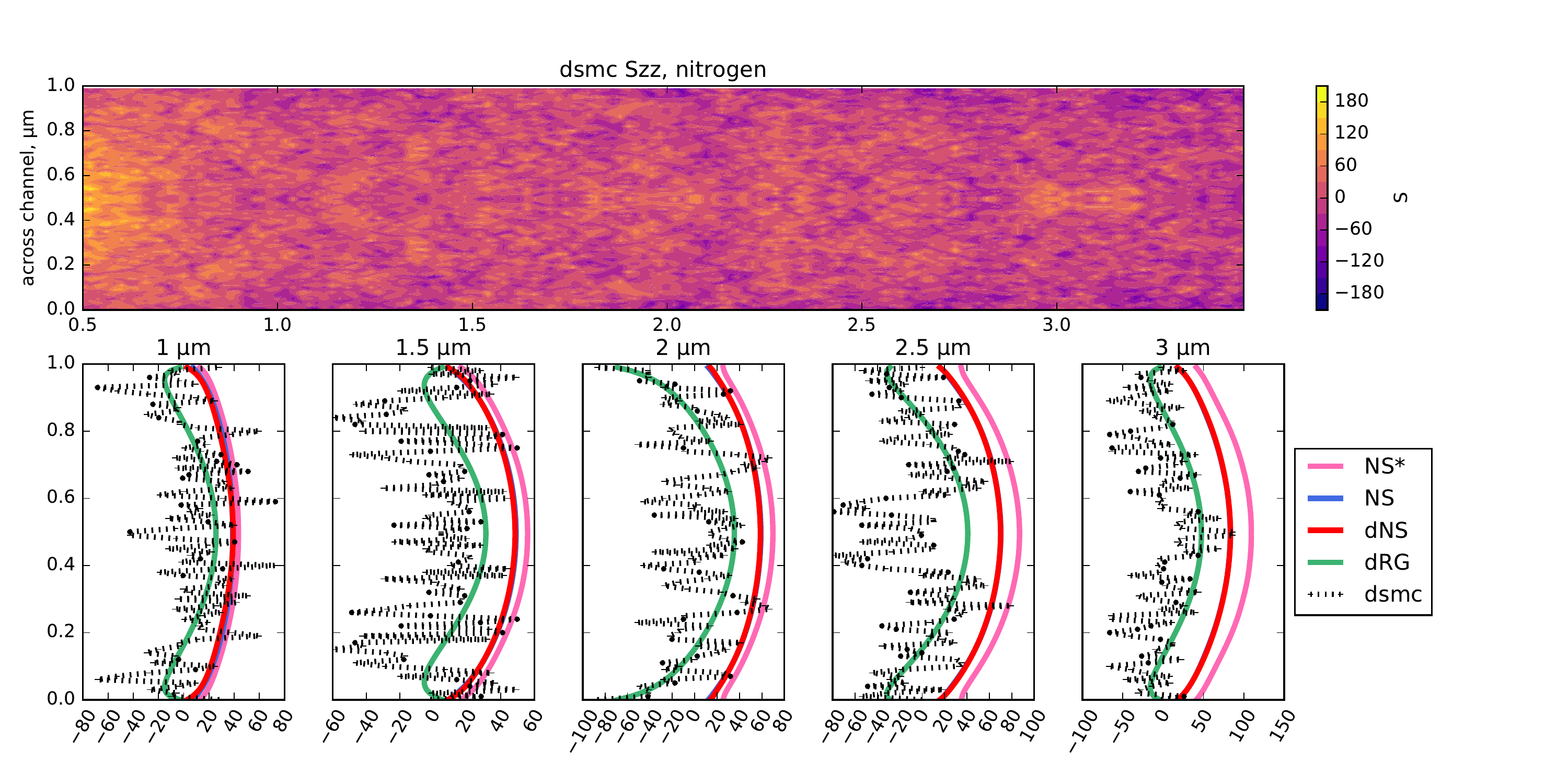}
\caption{Stress component $S_{zz}$ of the Poiseuille flow of
  nitrogen (m$^2$/s$^2$).}
\label{fig:nitrogen_Szz}
\end{figure}

\begin{figure}
\includegraphics[width=\textwidth]{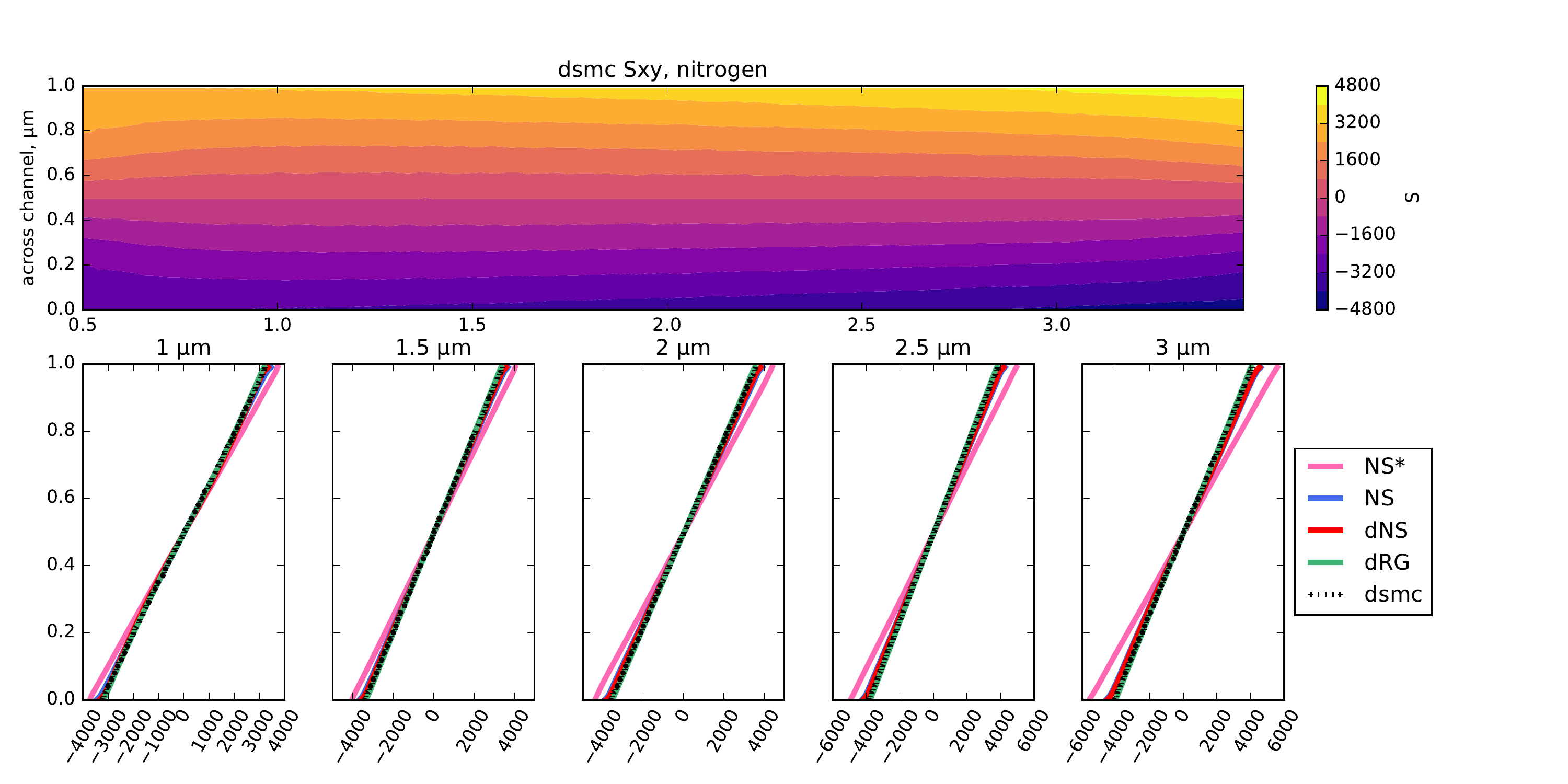}
\caption{Stress component $S_{xy}$ of the Poiseuille flow of
  nitrogen (m$^2$/s$^2$).}
\label{fig:nitrogen_Sxy}
\end{figure}

\begin{figure}
\includegraphics[width=\textwidth]{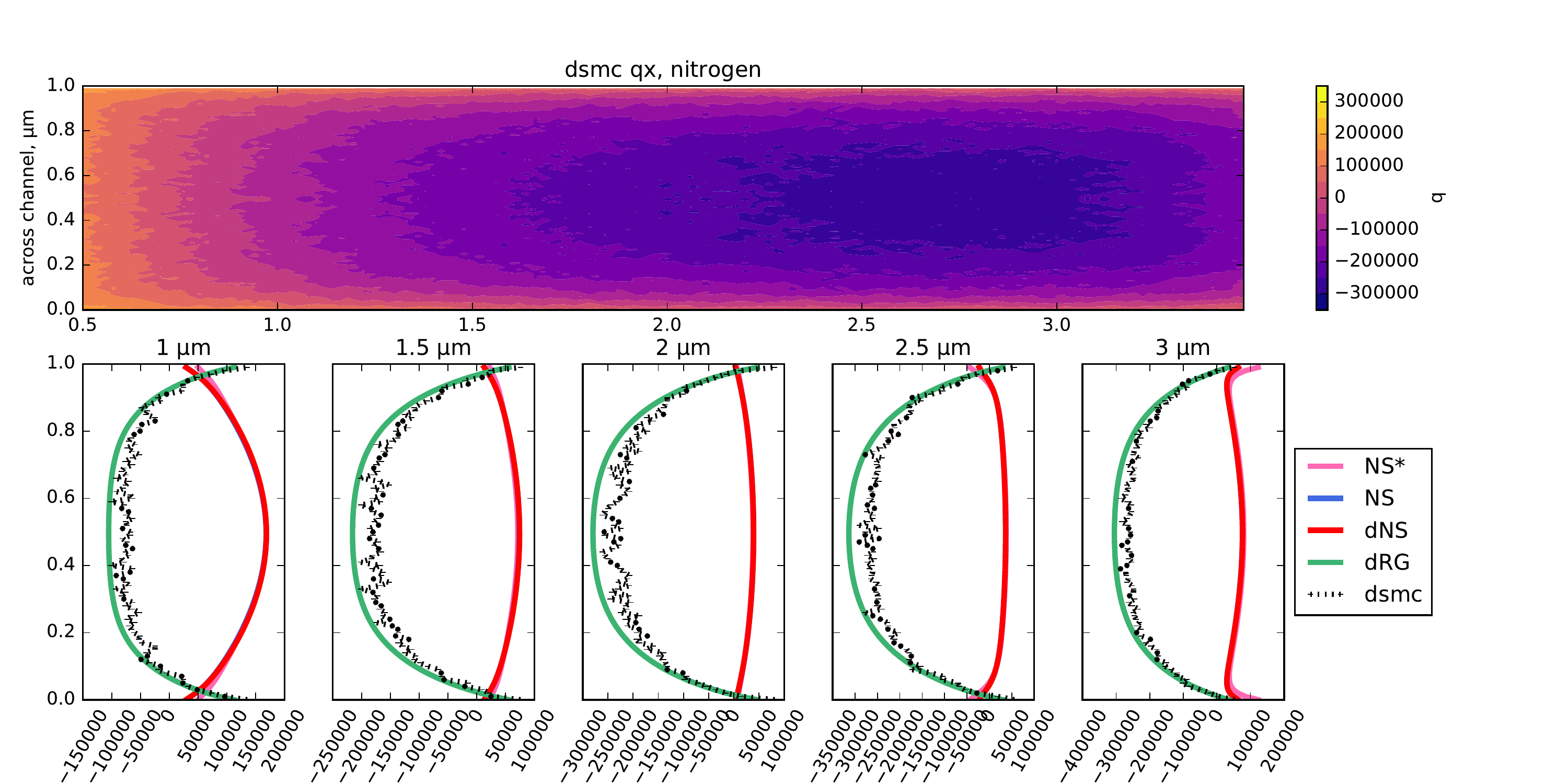}
\caption{Heat flux component $q_x$ of the Poiseuille flow of
  nitrogen (m$^3$/s$^3$).}
\label{fig:nitrogen_qx}
\end{figure}

\begin{figure}
\includegraphics[width=\textwidth]{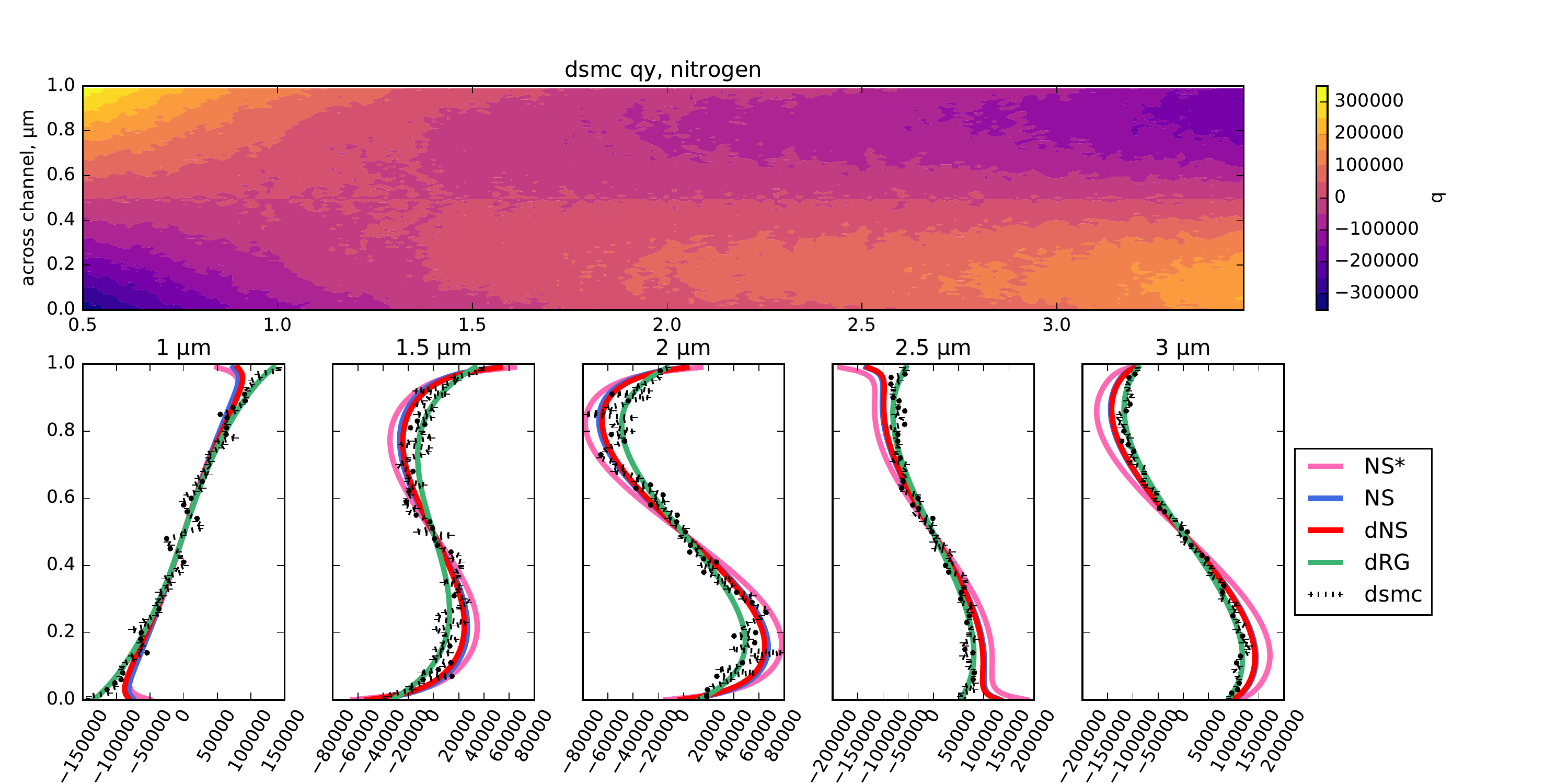}
\caption{Heat flux component $q_x$ of the Poiseuille flow of
  nitrogen (m$^3$/s$^3$).}
\label{fig:nitrogen_qy}
\end{figure}

Additionally, in Figures \ref{fig:argon_Sxx}--\ref{fig:argon_qy} we
show the components of the stress and heat flux for the continuum gas
dynamics closures and compare them against the DSMC computation.
Observe that the diffusive regularized Grad closure is the only
closure which provides moderately accurate approximations of the
higher-order moments. The Navier-Stokes approximations show a good
qualitative correspondence for the cross-stress component $S_{xx}$ and
the transversal heat flux component $q_y$, while all other stress and
heat flux components are qualitatively different from the DSMC
results.

\subsection{Nitrogen}

Nitrogen is a diatomic gas and a major component of Earth
atmosphere. We used the following computational parameters for argon:
\begin{itemize}
\item The adiabatic constant was set to $\gamma=1.4$,
\item The Prandtl number was set to $\Pran=0.69$,
\item The higher-order Prandtl numbers for the regularized Grad
  equations were set to the same values as for argon
  above~\cite{Abr13}, $\Pran_{\widetilde{\!\BM Q}}=3/2$,
  $\Pran_{\widetilde R}=2/3$ and $\Pran_{\widetilde{\!\BM R}}=7/6$,
\item The molar mass $M$ was set to $2.801\cdot 10^{-2}$ kg/mol,
\item The viscosity $\mu$ and empirically scaled mass diffusivity were
  as in~\eqref{eq:mu_D} above, with the reference temperature set to
  $T^*=288.15$ K, whereas the reference viscosity and empirical scaled
  mass diffusivity constants were set to $\mu^*=1.74\cdot 10^{-5}$
  kg/(m sec) \cite{HirCurBir,LemJac}, and $D_\alpha^*=10^{-6}$ kg/(m
  sec), respectively.
\end{itemize}
In Figures \ref{fig:nitrogen_Ux}--\ref{fig:nitrogen_errors_Tp} we show
the velocity, mass flow (the product of the density with velocity),
temperature and pressure for the continuum gas dynamics closures and
compare them against the DSMC computation. Just as above for argon,
observe that the conventional Navier-Stokes closure without the
viscosity scaling in~\eqref{eq:viscosity_scaling} is consistently the
least accurate continuum gas dynamics approximation for nitrogen among
all tested -- its relative error against the DSMC computation reaches
7-8\% in velocity, mass flow and pressure. On the other hand, the
diffusive regularized Grad closure with the viscosity scaling
in~\eqref{eq:viscosity_scaling} is consistently the most accurate
continuum gas dynamics closure -- its relative error against the DSMC
computation is about 1\% in velocity and mass flow, and about 0.2\% in
pressure. Just as above for argon, the temperature is approximated
well by all continuum gas dynamics closures -- the errors in
temperature are about 0.1\% irrespective of the closure.

Additionally, in Figures \ref{fig:nitrogen_Sxx}--\ref{fig:nitrogen_qy}
we show the components of the stress and heat flux for the continuum
gas dynamics closures and compare them against the DSMC computation.
Just as above for argon, observe that the diffusive regularized Grad
closure is the only continuum gas dynamics closure which provides
moderately accurate approximations of the higher-order moments. The
Navier-Stokes approximations show a good qualitative correspondence
for the cross-stress component $S_{xx}$ and the transversal heat flux
component $q_y$, while all other stress and heat flux components are
qualitatively different from the DSMC results.

\section{Summary}
\label{sec:summary}

We implement and test the nonequilibrium diffusive gas dynamics
equations~\cite{Abr13} with the near-wall viscosity
scaling~\cite{Abr15}, considering the Poiseuille microflow in a one
micrometer wide channel at close-to-normal thermodynamic conditions
(with temperature about $288$ K and pressure between one-half and
one-third of that at sea level), simulating both argon (monatomic) and
nitrogen (diatomic) gases. By comparison to the statistical DSMC
computation, of all the gas dynamics closures tested, the diffusive
regularized Grad equations~\cite{Abr13} with near-wall viscosity
scaling~\cite{Abr15} were consistently the most accurate. Its relative
errors in thermodynamic variables (i.e. velocity, mass flow,
temperature, and pressure) were about 1\% in the worst cases, and we
saw qualitatively correct representation of the higher-order moments
(stress and heat flux). By contrast, the conventional Navier-Stokes
equations without the viscosity scaling were consistently the least
accurate (with relative errors up to 10\%), and the resulting
parameterizations for the stress and heat flux qualitatively differed
from those recovered by DSMC.

\ack The first author was supported by the Office of Naval Research
grant N00014-15-1-2036. The second author was supported as a Research
Assistant by the same grant.

\end{document}